\documentclass{aastex}
\usepackage{spr-astr-addons}             

\RequirePackage{color}                   

\begin{document}

\title{CCD {\it UBV} photometric study of five open clusters - Dolidze\,36, NGC\,6728, NGC\,6800, NGC\,7209, and Platais\,1}
\slugcomment{Not to appear in Nonlearned J., 45.}
\shorttitle{CCD {\it UBV} photometric study of five open clusters}
\shortauthors{Z.~F. Bostanc\i, T. Yontan, S. Bilir, T. Ak, T. G\"uver, S. Ak, E. Paunzen, \c C. S. Ba\c saran, E. Vurgun, A. B. Akti, M. \c Celebi, and H. \"Urg\"up}

\author{Z.~F. Bostanc\i \altaffilmark{1}}
\altaffiltext{1}{Istanbul University, Faculty of Science, Department 
of Astronomy and Space Sciences, 34119, University, Istanbul, Turkey\\
\email{funda.bostanci@istanbul.edu.tr}}

\author{T. Yontan \altaffilmark{2}}
\altaffiltext{2}{Istanbul University, Institute of Graduate 
Studies in Science and Engineering, Programme of Astronomy and 
Space Sciences, 34116, Beyaz{\i}t, Istanbul, Turkey\\}

\author{S. Bilir \altaffilmark{1}}
\altaffiltext{1}{Istanbul University, Faculty of Science, Department 
of Astronomy and Space Sciences, 34119, University, Istanbul, Turkey\\}

\author{T. Ak \altaffilmark{1}}
\altaffiltext{1}{Istanbul University, Faculty of Science, Department 
of Astronomy and Space Sciences, 34119, University, Istanbul, Turkey\\}

\author{T. G\"uver \altaffilmark{1,3}}
\altaffiltext{1}{Istanbul University, Faculty of Science, Department 
of Astronomy and Space Sciences, 34119 University, Istanbul, Turkey\\}
\altaffiltext{3}{Istanbul University Observatory Research and Application 
Center, Beyaz{\i}t, 34119, Istanbul, Turkey\\}

\author{S. Ak \altaffilmark{1}}
\altaffiltext{1}{Istanbul University, Faculty of Science, Department 
of Astronomy and Space Sciences, 34119, University, Istanbul, Turkey\\}

\author{E. Paunzen \altaffilmark{4}}
\altaffiltext{4}{Department of Theoretical Physics and Astrophysics, 
Masaryk University, Kotl\'a\u rsk\'a 2, 611 37 Brno, Czech Republic\\}

\author{\c C. S. Ba\c saran \altaffilmark{2}}
\altaffiltext{2}{Istanbul University, Institute of Graduate 
Studies in Science and Engineering, Programme of Astronomy and 
Space Sciences, 34116, Beyaz{\i}t, Istanbul, Turkey\\}

\author{E. Vurgun \altaffilmark{2}}
\altaffiltext{2}{Istanbul University, Institute of Graduate 
Studies in Science and Engineering, Programme of Astronomy and 
Space Sciences, 34116, Beyaz{\i}t, Istanbul, Turkey\\}

\author{B. A. Akti \altaffilmark{2}}
\altaffiltext{2}{Istanbul University, Institute of Graduate 
Studies in Science and Engineering, Programme of Astronomy and 
Space Sciences, 34116, Beyaz{\i}t, Istanbul, Turkey\\}

\author{M. \c Celebi \altaffilmark{2}}
\altaffiltext{2}{Istanbul University, Institute of Graduate 
Studies in Science and Engineering, Programme of Astronomy and 
Space Sciences, 34116, Beyaz{\i}t, Istanbul, Turkey\\}

\and
\author{H. \"Urg\"up \altaffilmark{2}}
\altaffiltext{2}{Istanbul University, Institute of Graduate 
Studies in Science and Engineering, Programme of Astronomy and 
Space Sciences, 34116, Beyaz{\i}t, Istanbul, Turkey\\}

\begin{abstract} In this study, we present CCD {\it UBV} photometry of
poorly studied open star clusters, Dolidze\,36, NGC\,6728, NGC\,6800, NGC\,7209,
and Platais\,1, located in the first and second Galactic quadrants. 
Observations were obtained with T100, the 1-m telescope of the T\"UB\.ITAK National
Observatory. Using  photometric data, we determined several astrophysical
parameters such as reddening, distance, metallicity and ages and from them,
initial mass functions, integrated magnitudes and colours.
We took into account the proper motions of the observed stars to  calculate
the membership probabilities. The colour excesses and metallicities were
determined independently using two-colour diagrams. After obtaining the
colour excesses of the clusters Dolidze\,36, NGC\,6728, NGC\,6800, NGC\,7209, and
Platais\,1 as $0.19\pm0.06$, $0.15\pm0.05$, $0.32\pm0.05$, $0.12\pm0.04$, and
$0.43\pm0.06$ mag, respectively, the metallicities are found to be
$0.00\pm0.09$, $0.02\pm0.11$, $0.03\pm0.07$, $0.01\pm0.08$, and $0.01\pm0.08$
dex, respectively. Furthermore, using these parameters, distance moduli and
age of the clusters were also calculated from colour-magnitude diagrams
simultaneously using PARSEC theoretical models. The distances to the clusters
Dolidze\,36, NGC\,6728, NGC\,6800, NGC\,7209, and Platais\,1 are  $1050\pm90$,
$1610\pm190$, $1210\pm150$, $1060\pm 90$, and $1710\pm250$ pc, respectively,
while corresponding ages are $400\pm100$, $750\pm150$, $400\pm100$,
$600\pm100$, and $175\pm50$ Myr, respectively. Our results are compatible
with those found in previous studies. The mass function of each cluster is derived. 
The slopes of the mass functions of the open clusters range from
1.31 to 1.58, which are in agreement with Salpeter's initial mass function.
We also found integrated absolute magnitudes varying from -4.08 to -3.40 for
the clusters. \end{abstract}



\section{Introduction} Open star clusters are important tools to study
Galactic chemical composition and structure, dynamical evolution and star
formation processes in the Galaxy. Since their members are formed within a
few millions years simultaneously from the same molecular cloud, they are
almost at the same age and distance with comparable chemical composition, 
but different stellar masses. Hence, open clusters give us great opportunity to determine
structural and astrophysical parameters of a group of stars such as
reddening, distance, metallicity, age, and then mass function, integrated
magnitudes and colours.

In this context, we studied CCD {\it UBV} stellar photometry of five open
clusters (Dolidze\,36, NGC\,6728, NGC\,6800, NGC\,7209, and Platais\,1, 
see Table 1) located in the first and second Galactic quadrants and investigated
their basic astrophysical parameters (reddening, distance, metallicity and
age) as well as initial mass functions, integrated magnitudes and colours in
detail. We used the technique that is based on the analysis of the two-colour
diagrams (TCDs) and colour-magnitude diagrams (CMDs) of member stars of the
clusters \citep[cf.][]{Bica06, Yontan15}

CCD {\it UBV} photometry of most of the clusters in our sample have not
been examined closely in previous studies. We present physical parameters of
the five open clusters inferred in previous studies from the literature to
date in Table 1. Our paper is organized as follows. We briefly define the
observations and reductions in Section 2. We then give the CMDs of the five
open clusters and the membership probabilities of the stars in the respective
fields in Section 3. We obtain the astrophysical parameters of each cluster
in Section 4 and summarize our conclusions in Section 5. 

\begin{table*}[t]
\setlength{\tabcolsep}{6pt}
{\scriptsize
\begin{center}
\caption{Galactic coordinates ($l$, $b$), colour excesses ($E(B-V)$), distance moduli
($\mu_V$), distances ($d$), iron abundance ([Fe/H]), and age ($t$) collected from the
literature for five open clusters under investigation. The references are given in the
last column.}
\begin{tabular}{lcccccccr}
\hline
\multicolumn{1}{c}{Cluster} & $l$    & $b$    & $E(B-V)$ & $\mu_V$ & $d$  & [Fe/H] & $\log t$ & \multicolumn{1}{c}{References}\\
          & $(^o)$ & $(^o)$ & (mag)    & (mag) & (pc) & (dex)  &          &            \\
\hline
Dolidze~36& 77.66 & +5.98  & 0.22          & 10.45          & 900         & --             & 8.83         & \citet{Kharchenko05}  \\
          &       &        & 0.19$\pm$0.06 & 10.70$\pm$0.19 & 1050$\pm$90 & 0.00$\pm$0.09  & 8.60$\pm$0.10& This study            \\
NGC\,6728  & 25.76 &$-$5.70 & 0.15          & 10.47          & 1000        & --             & 8.93         & \citet{Kharchenko05}  \\
          &       &        & 0.15$\pm$0.05 & 11.50$\pm$0.25 & 1610$\pm$190& 0.02$\pm$0.11  & 8.88$\pm$0.08& This study            \\
NGC\,6800  & 59.23 & +3.95  & 0.40          & 11.29          & 1025        & --             & 8.40         & \citet{Ananjevskaja15}\\
          &       &        & 0.40          & 11.24          & 1000        & --             & 8.59         & \citet{Kharchenko05}  \\
          &       &        & 0.32$\pm$0.05 & 11.40$\pm$0.26 & 1210$\pm$150& 0.03$\pm$0.07  & 8.60$\pm$0.10& This study            \\
NGC\,7209  & 95.50 &$-$7.34 & 0.16          & 10.83          & 1167        & --             & 8.65         & \citet{Kharchenko05}  \\
          &       &        & 0.171         & 10.60$\pm$0.06 & 1026$\pm$30 & -0.07$\pm$0.03 & 8.65         & \citet{Vansevicius97} \\
          &       &        & 0.173         & ---            & 1095        & --             & 8.48         & \citet{Malysheva97}   \\
          &       &        & 0.17          & 10.75          & 1140        & 0.07$\pm$0.05  & --           & \citet{Twarog97}      \\
          &       &        & 0.15$\pm$0.04 & 10.25          & 900         & -0.01$\pm$0.11 & 8.48         & \citet{Claria96, Lynga87}\\
          &       &        & 0.15          & --             & 900         & -0.12          & 8.48         & \citet{Piatti95, Lynga87}\\
          &       &        & 0.12$\pm$0.04 & 10.50$\pm$0.18 & 1060$\pm$90 & 0.01$\pm$0.08  & 8.78$\pm$0.07& This study               \\
Platais~1 & 92.56 &$-$1.65 & 0.43$\pm$0.02 & 12.30$\pm$0.02 & 1568$\pm$13 & --             & 8.04-8.28    & \citet{Turner94}         \\
          &       &        & 0.43$\pm$0.06& 12.50$\pm$0.29 & 1710$\pm$250 & 0.01$\pm$0.08  & 8.24$\pm$0.11& This study               \\
\hline
\end{tabular}
\end{center}
}
\end{table*}

\begin{table*}
\setlength{\tabcolsep}{5pt}
\begin{center}
\caption{Log of observations with dates and exposure times for each passband. $N$ refers to the number of exposure.}
\begin{tabular}{lcccc}
\hline
 & & \multicolumn{3}{c}{Filter / (Exposure Time (s) $\times N$)}\\
Cluster & Observation Date & $U$ & $B$ & $V$ \\
\hline
Dolidze~36 & 17.08.2012 & 90$\times$3, 360$\times$3 & 5$\times$3, 60$\times$3 & 3$\times$3, 30$\times$4 \\
NGC~6728   & 19.07.2012 & 3$\times$3, 60$\times$1, 120$\times$2 & 0.6$\times$3, 1$\times$1, 25$\times$4 & 0.25$\times$3, 7$\times$3 \\
NGC~6800   & 20.07.2012 & 90$\times$3, 360$\times$3 & 5$\times$3, 60$\times$3 & 3$\times$3, 30$\times$2 \\
NGC~7209   & 16.08.2012 & 60$\times$3, 360$\times$3 & 5$\times$3, 60$\times$3 & 2$\times$3, 20$\times$3 \\
Platais~1  & 08.08.2013 & 60$\times$3  & 4$\times$3  & 2$\times$4 \\
\hline
\end{tabular}
\end{center}
\end{table*}

\begin{figure}
\includegraphics[scale=0.50, angle=0]{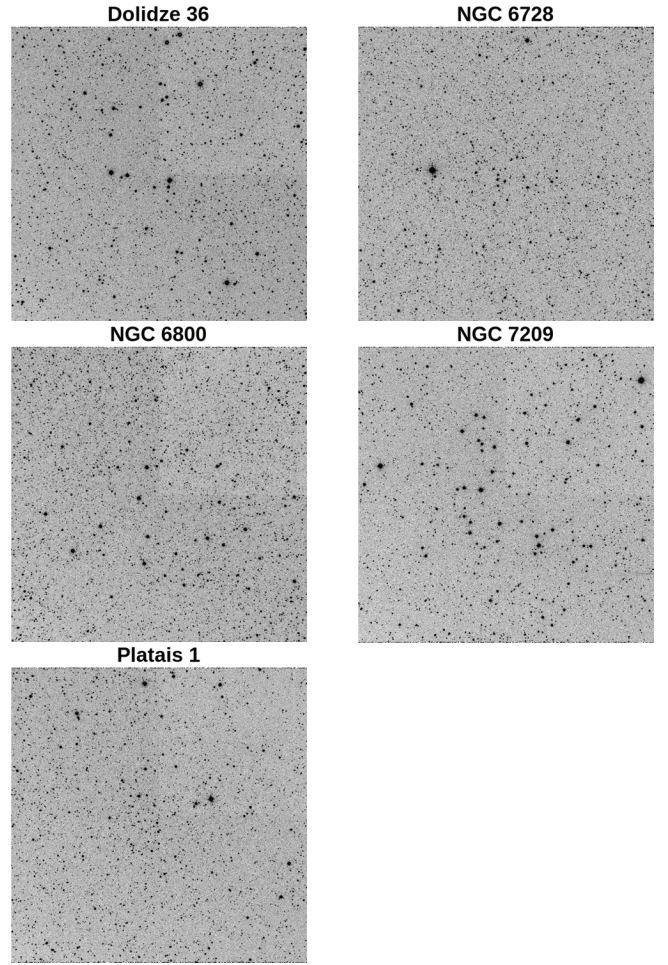}
\caption{Inverse coloured $V$-band images of five open clusters taken
with T100 telescope. The field of view is about 21$\times$21 arcmin (North
top and East left).} 
\end {figure}

\begin{figure}[!htb]
\centering
\includegraphics[scale=0.72, angle=0]{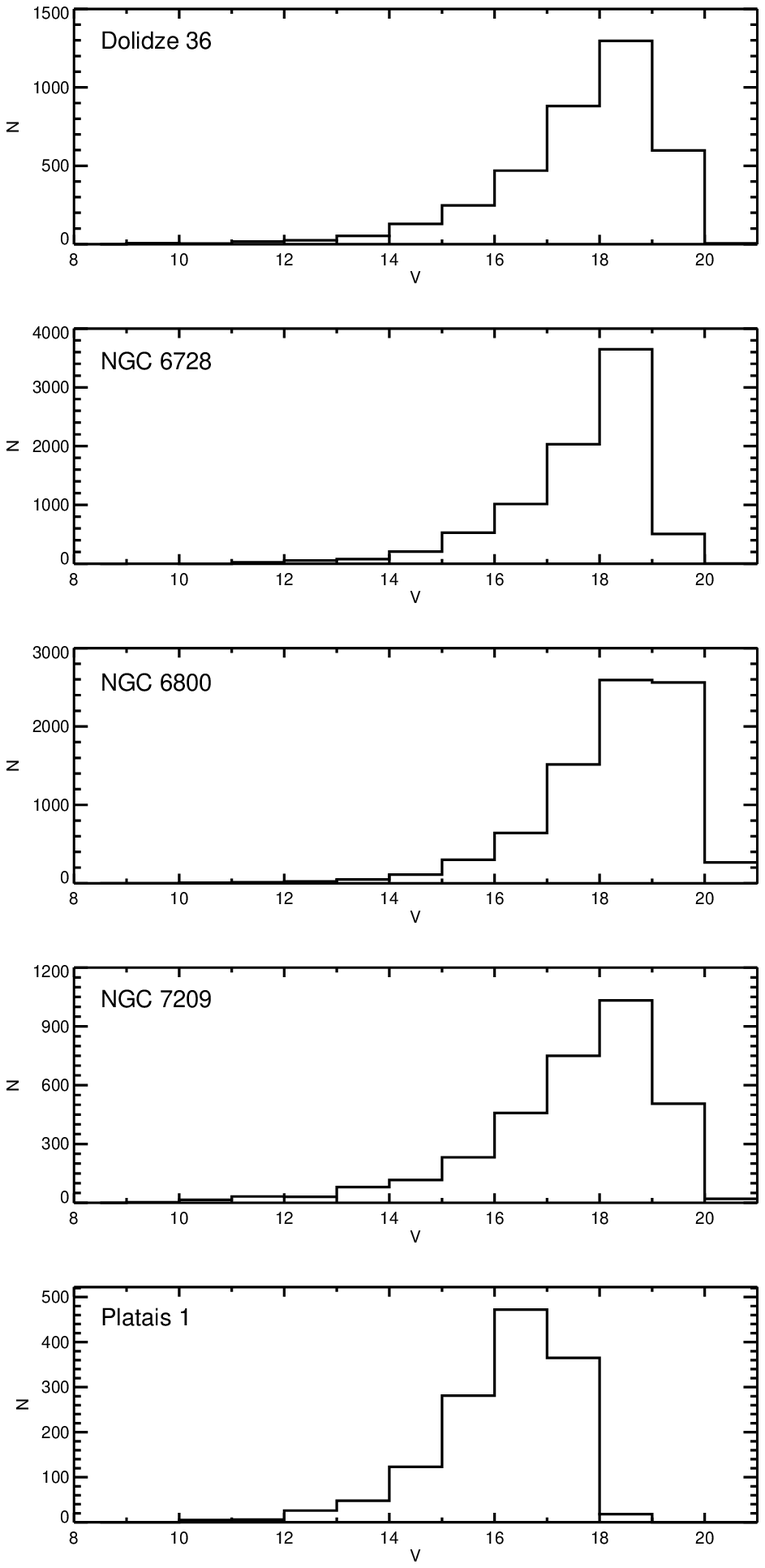}
\caption{Histograms of the $V$-band magnitudes measured for five
clusters. Photometric completeness limits of the data are 19 mag for Dolidze~36,
NGC\,6728, NGC\,6800, and NGC\,7209, and 17 mag for Platais\,1.} 
\end {figure}


\section{Observations}

CCD {\it UBV}  observations of the five open clusters were performed with the
1-m Ritchey-Chr\'etien telescope (T100) of the T\"UB\.ITAK  National
Observatory (TUG)\footnote{www.tug.tubitak.gov.tr} located in Turkey. Inverse
coloured $V$-band images of the clusters are given in Fig.\,1. Each frame
contains an integrated image of the largest exposure obtained.

Cluster images were taken using a Spectral Instruments (SI 1100) CCD camera
operating at $-90^o \rm{C}$. The camera is equipped with a back illuminated
4k$\times$4k pixels CCD,  with a pixel scale of $0.''31$ pixel$^{-1}$,
resulting in an unvignetted field of view of about $21' \times 21'$. The
readout noise and the gain of the CCD camera is 4.19~e$^{-}$ and 0.55
e$^{-}$/ADU, respectively. In order to be sensitive to the widest possible
flux range, different exposure times were used during the observations. A log
of observations is given in Table 2. Due to poor seeing it was not possible
to use long exposure data of Platais\,1.
IRAF\footnote{IRAF is distributed by the
National Optical Astronomy Observatories}, PyRAF\footnote{PyRAF is a product
of the Space Telescope Science Institute, which is operated by AURA for NASA}
and astrometry.net\footnote{http://astrometry.net} \citep{Lang10} routines
were used together with our own scripts for CCD calibrations as well as the
astrometric corrections of the images. 
In order to determine atmospheric extinction and transformation coefficients for each night, 
we observed several star fields containing more than one \citet{Landolt09}'s standard stars.
We took three images per band for each field and some fields could be observed several times. 
Details of the observations are given in Table 3. Instrumental magnitudes of the
standard stars were measured with the aperture photometry packages of IRAF
while brightness of the objects in cluster fields were obtained with Source
Extractor and PSF Extractor routines \citep{Bertin96}. Applying multiple linear fits
to the instrumental magnitudes of the standard stars, we obtained the photometric
extinction and transformation coefficients for each night. We list the extinction
coefficients and transformation coefficients for the individual nights in Table 4.
After the application of aperture correction to instrumental magnitudes, 
stellar magnitudes were obtained with transformation equations given in previous 
studies \citep{Janes2013, Yontan15}.

\begin{table*}[t]
\scriptsize
\setlength{\tabcolsep}{5pt}
\centering
\caption{Observations of \citet{Landolt09}'s standard star fields, 
Field name (Field), number of stars ($N_{st}$), number of pointings 
to the fields ($N_{obs}$), airmass range for all fields ($X_{range}$) 
are given.}
\begin{tabular}{ccccc}
\hline
Date & Field & $N_{st}$ & $N_{obs}$ & $X_{range}$ \\
\hline
               & PG1530+057 & 3 & 1 &  \\
               & PG2213-006 & 7 & 1 &  \\
               & SA110 & 10 & 1 &  \\
19.07.2012     & SA111 & 5  & 1 & 1.184 - 2.571 \\
               & SA112 & 6 & 10 &  \\
               & SA113 & 15 & 1 &  \\
               & SA114 & 5 & 1 &  \\
\hline
20.07.2012     & PG1530+057 & 3 & 1 & 1.183 - 2.612\\
               & SA110 & 10 & 1 &  \\
               & SA111 & 5 & 1 &  \\
               & SA112 & 6 & 5 &  \\
\hline
               & F24 & 4 & 1 &  \\
               & G93 & 5 & 11 &  \\
               & PG1633+099 & 8 & 1 &  \\
               & PG2213-006 & 7 & 1 &  \\
16.08.2012     & SA108 & 2 & 1 &  1.164 - 3.196\\
               & SA110 & 10 & 1 &  \\
               & SA112 & 6 & 1 &  \\
               & SA92 & 6 & 1 &  \\
               & SA93 & 4 & 1 &  \\
\hline
               & F24 & 4 & 1 & \\
               & G93 & 5 & 15 &  \\
               & PG1633+099 & 8 & 1 &  \\
               & PG2213-006 & 7 & 1 &  \\
17.08.2012     & SA108 & 2 & 1 & 1.182 - 2.889 \\
               & SA110 & 10 & 1 &  \\
               & SA112 & 6 & 1 &  \\
               & SA92 & 6 & 1 &  \\
               & SA93 & 4 & 1 &  \\
\hline
               & F11        & 3  & 1 & \\
               & G93        & 5  & 2 & \\	
               & GD246      & 4	 & 1 & \\	
               & PG2213-006 & 7	 & 1 & \\	
08.08.2013     & SA107      & 7	 & 1 & 1.153 - 2.466\\	
               & SA111      & 5	 & 1 & \\	
               & SA112      & 6	 & 6 & \\	
               & SA113      & 15 & 1 & \\	
               & SA114      & 5	 & 1 & \\	
               & SA92       & 21 & 1 & \\	
\hline
\end{tabular}
\end{table*}


\begin{table*}[!htb]
\scriptsize
\setlength{\tabcolsep}{5pt}
\centering
\caption{Derived transformation and extinction coefficients. $k$ and $k^{'}$ are primary
and secondary extinction coefficients, respectively, while $\alpha$ and $C$ are transformation
coefficients.}
\begin{tabular}{ccccc}
\hline
Obs. Time & \multicolumn{4}{c}{19.07.2012}\\
\hline
Indice & $k$ & $k^{'}$ & $\alpha$ & $C$ \\
\hline
$U$ & 0.472 $\pm$ 0.031 & -0.019 $\pm$ 0.033 & -- & -- \\
$B$ & 0.326 $\pm$ 0.024 & -0.057 $\pm$ 0.025 & 0.992 $\pm$ 0.038 & 0.745 $\pm$ 0.036 \\
$V$ & 0.189 $\pm$ 0.007 & -- & -- & -- \\
$U-B$ & -- & -- & 0.861 $\pm$ 0.050 & 3.143 $\pm$ 0.046 \\
$B-V$ & -- & -- & 0.077 $\pm$ 0.011 & 0.799 $\pm$ 0.017 \\
\hline
Obs. Time & \multicolumn{4}{c}{20.07.2012}\\
\hline
$U$ & 0.559 $\pm$ 0.033 & -0.002 $\pm$ 0.035 & -- & -- \\
$B$ & 0.392 $\pm$ 0.011 & -0.037 $\pm$ 0.011 & 0.936 $\pm$ 0.018 & 0.786 $\pm$ 0.017 \\
$V$ & 0.235 $\pm$ 0.002 & -- & -- & -- \\
$U-B$ & -- & -- & 0.813 $\pm$ 0.056 & 3.190 $\pm$ 0.054 \\
$B-V$ & -- & -- & 0.051 $\pm$ 0.003 & 0.838 $\pm$ 0.005 \\
\hline
Obs. Time & \multicolumn{4}{c}{16.08.2012} \\
\hline
$U$ & 0.678 $\pm$ 0.030 & -0.023 $\pm$ 0.067 & -- & -- \\
$B$ & 0.525 $\pm$ 0.050 & -0.246 $\pm$ 0.079 & 1.254 $\pm$ 0.105 & 0.550 $\pm$ 0.067 \\
$V$ & 0.242 $\pm$ 0.007 & -- & -- & -- \\
$U-B$ & -- & -- & 0.836 $\pm$ 0.093 & 3.041 $\pm$ 0.043 \\
$B-V$ & -- & -- & 0.084 $\pm$ 0.011 & 0.779 $\pm$ 0.016 \\
\hline
Obs. Time & \multicolumn{4}{c}{17.08.2012} \\
\hline
$U$ & 0.365 $\pm$ 0.026 & 0.216 $\pm$ 0.041 & -- & -- \\
$B$ & 0.270 $\pm$ 0.041 & 0.002 $\pm$ 0.051 & 0.918 $\pm$ 0.066 & 0.890 $\pm$ 0.054 \\
$V$ & 0.166 $\pm$ 0.004 & -- & -- & -- \\
$U-B$ & -- & -- & 0.537 $\pm$ 0.055 & 3.453 $\pm$ 0.034 \\
$B-V$ & -- & -- & 0.081 $\pm$ 0.011 & 0.879 $\pm$ 0.015 \\
\hline
Obs. Time & \multicolumn{4}{c}{08.08.2013} \\
\hline
$U$ & 0.397 $\pm$ 0.014 & 0.007 $\pm$ 0.018 & -- & -- \\
$B$ & 0.240 $\pm$ 0.014 & -0.041 $\pm$ 0.015 & 0.972 $\pm$ 0.024 & 0.481 $\pm$ 0.021 \\
$V$ & 0.118 $\pm$ 0.003 & -- & -- & -- \\
$U-B$ & -- & -- & 0.840 $\pm$ 0.027 & 2.893 $\pm$ 0.020 \\
$B-V$ & -- & -- & 0.070 $\pm$ 0.003 & 0.538 $\pm$ 0.005 \\
\hline
\end{tabular}
\end{table*}

\section{Data Analysis} \subsection{Photometry of the detected objects}
We constructed catalogues including all the sources in the field of view of
five open clusters. All photometric catalogues of the open clusters are given
electronically. In these catalogues, we present equatorial coordinates,
apparent magnitude ($V$), colours ($U-B$, $B-V$), proper motion components
\citep[$\mu_{\alpha}\cos\delta$, $\mu_{\delta}$;][]{Roeser10} and the
probability of membership ($P$), respectively.

We listed mean errors of the measurements in the $V$ band, and $U-B$ and
$B-V$ colours in the selected apparent $V$ magnitude ranges in Table 5.
Errors at brighter ranges (mostly $V<17$ mag) are relatively small however,
they increase exponentially towards fainter objects. To calculate the precise
astrophysical parameters of the clusters, it is important to know the
photometric completeness limit of the data. In order to determine this limit,
we composed histograms of $V$ magnitudes for each cluster (given in Fig.\,2).
Modes of the distribution of $V$ magnitudes vary from 17 to 19 for the
clusters used in this study. Platais\,1 has a brighter limit of 17 mag since 
we took into account only observations with short exposure times.  
Only the calculations including stars brighter
than the found completeness limits will be able to offer reliable results in
the determination of the astrophysical parameters of the clusters.

\begin{table*}
\setlength{\tabcolsep}{2pt}
{\scriptsize
\begin{center}
\caption{Mean errors of the photometric measurements for the stars in the
directions of five open clusters. $N$ represents the number of stars within
the $V$ apparent magnitude range given in the first column.}
\begin{tabular}{rcccccccccccccccccccc}
\hline
 & \multicolumn{4}{c}{Dolidze~36} & \multicolumn{4}{c}{NGC~6728} & \multicolumn{4}{c}{NGC~6800} & \multicolumn{4}{c}{NGC~7209}  & \multicolumn{4}{c}{Platais~1}\\
\hline
$V$ range & $N$ & $\sigma_V$ & $\sigma_{U-B}$ & $\sigma_{B-V}$ & $N$ & $\sigma_V$ & $\sigma_{U-B}$ & $\sigma_{B-V}$ & $N$ & $\sigma_V$ & $\sigma_{U-B}$ & $\sigma_{B-V}$ & $N$ & $\sigma_V$ & $\sigma_{U-B}$ & $\sigma_{B-V}$ & $N$ & $\sigma_V$ & $\sigma_{U-B}$ & $\sigma_{B-V}$\\
\hline
(8, 10]  & 7	& 0.001 & 0.002 & 0.001& 1    & 0.002 & 0.002 & 0.002 & ---  & ---   & ---   & ---   & 3    & 0.001 & 0.002 & 0.001  & --- & ---   & ---   & ---   \\
(10, 12] & 21	& 0.002 & 0.003 & 0.002& 24   & 0.007 & 0.022 & 0.009 & 19   & 0.002 & 0.002 & 0.002 & 47   & 0.002 & 0.002 & 0.002  & 11  & 0.002 & 0.003 & 0.002 \\
(12, 14] & 78	& 0.004 & 0.014 & 0.008& 133  & 0.013 & 0.050 & 0.016 & 72   & 0.004 & 0.011 & 0.007 & 110  & 0.004 & 0.007 & 0.005  & 74  & 0.006 & 0.018 & 0.010 \\
(14, 16] & 377  & 0.003 & 0.010 & 0.005& 735  & 0.006 & 0.029 & 0.008 & 409  & 0.004 & 0.014 & 0.006 & 348  & 0.004 & 0.007 & 0.005  & 404 & 0.014 & 0.036 & 0.024 \\
(16, 18] & 1351 & 0.009 & 0.039 & 0.016& 3048 & 0.021 & 0.093 & 0.026 & 2157 & 0.014 & 0.045 & 0.020 & 1208 & 0.012 & 0.023 & 0.016  & 837 & 0.043 & 0.097 & 0.090 \\
(18, 20] & 1895 & 0.029 & 0.108 & 0.054& 4155 & 0.054 & 0.192 & 0.067 & 5153 & 0.048 & 0.122 & 0.074 & 1539 & 0.036 & 0.075 & 0.049  & 18  & 0.107 & 0.210 & 0.221 \\
(20, 22] & 5	& 0.076 & ---	& 0.152& 3    & 0.186 & 0.184 & 0.202 & 268  & 0.117 & 0.224 & 0.181 & 20   & 0.095 & 0.289 & 0.144  & --- & ---   & ---   & ---   \\
\hline
\end{tabular}
\end{center}
}
\end{table*}

Dolidze\,36 was observed within the scope of the Sloan Digital Sky Survey \citep[SDSS;][]{York00}. 
In order to figure out reliability of our photometry we compared {\it UBV}
observations with those calculated from SDSS. A cross-match of our catalogue of Dolidze\,36
with SDSS Data Release 12 \citep[DR12, ][]{Alam15} resulted in 370 stars. $V$ magnitudes of these common 
stars  are ranging between 14 and 19 because SDSS photometry aims towards faint stars. We used 
relations of \citet{Chonis08} to transform {\it ugriz} magnitudes to {\it UBV}. We show the comparison of observational and
calculated $V$ magnitude and $U-B$ and $B-V$ colours in Fig.\,3. We obtained means and
standard deviations of the magnitude and the colours from the differences as $\Delta_{V}$\,=\,0.088, 
$\sigma_{V}$\,=\,0.086, $\Delta_{U-B}$\,=\,0.140, $\sigma_{U-B}$\,=\,0.191 and $\Delta_{B-V}$\,=\,--0.048, 
$\sigma_{B-V}$\,=\,0.079 mag. It is seen that zero points between the two systems for $V$ and $B-V$ are 
usually small with very small standard deviations while they are somewhat larger for $U-B$. This 
large scattering in $U-B$ may arise from errors in the transformation since it is considered 
that the equations of \citet{Chonis08} do not take into account population types of the stars \citep{Bilir08, Bilir11}.

\begin{figure*}
\centering
\includegraphics[width=16.12cm, height=6.475cm, angle=0]{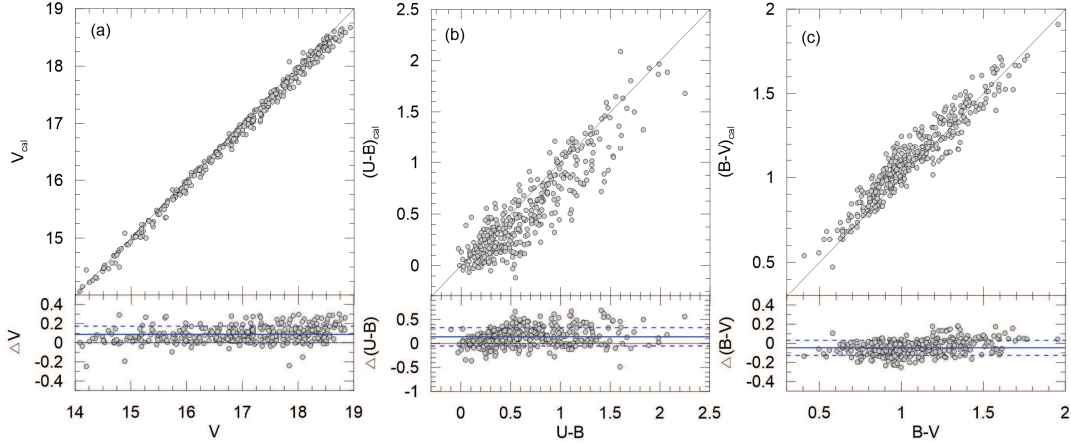}
\caption{Comparisons of observational magnitudes and colours with those calculated 
from SDSS DR12 \citep{Alam15}.} 
\end {figure*}

\subsection{Cluster radius and radial stellar surface density}
Since the studied open clusters show no noticeable central concentration except
Platais\,1 with little central concentration, we were only able to estimate
the stellar density profile for Platais\,1 using stars with
magnitudes brighter than the photometric completeness limit, $V = 17$ mag. The central
coordinates of the cluster were assumed to be as given in the WEBDA database\footnote{http://webda.physics.muni.cz}
($\alpha_{2000.0}=21^{h}30^{m}02^{s}$, $\delta_{2000.0}=+48^{\circ}58'36''$). 
The stellar density values for the main-sequence sample of the cluster have been 
evaluated in 1 arcminute steps. The last two annuli had a width of 2 arcmin because 
of a significant decrease in the number of stars (see, Fig.\,4).

\begin{figure}
\centering
\includegraphics[scale=0.40, angle=0]{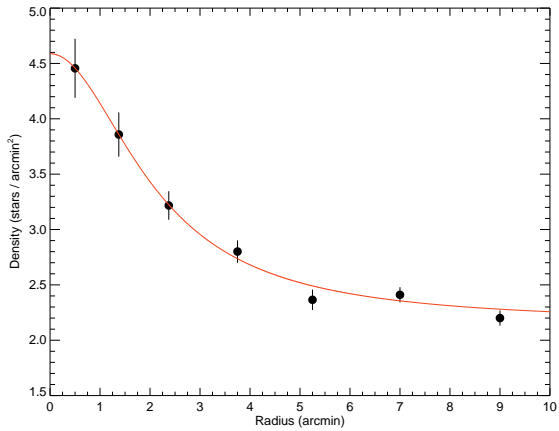}
\caption{Stellar density profile of Platais\,1. Errors were determined from sampling 
statistics, $\sqrt{N}$, where $N$ is the number of stars used in the density estimation.} 
\end {figure}

We fitted the \citet{King62} model to the observed radial density
profile and used a $\chi^2$ minimization technique to determine structural parameters.
The best fit to the density profile is shown with a solid line in Fig.~4. 
We found the central stellar density, core radius of the cluster, and the background 
stellar density as $f_0=2.43\pm0.25$ stars arcmin$^{-2}$, 
$r_c=2.10\pm0.75$ arcmin and $f_{bg}=2.16 \pm 0.27$ stars arcmin$^{-2}$,
respectively.

\citet{Turner94} estimated the stellar density profile of Platais\,1 for the limiting
$R$-band magnitude 17, using a photographic enlargement of the Palomar Observatory 
Sky Survey (POSS) E-plate. They only reported the background stellar density as 
$f_{bg}=4.38 \pm 0.13$ stars arcmin$^{-2}$. To infer their central stellar density and 
core radius of the cluster, we digitized values of their density profile from their figure 
and then fitted them with the King model \citep{King62}. As a result, the central stellar 
density and the core radius of the cluster were found as $f_0=3.25$ stars arcmin$^{-2}$ and 
$r_c=3.20$ arcmin, respectively. The core radius of the cluster is in agreement with the 
one in \citet{Turner94} within the quoted error, but the other parameters are a little 
different from each other. The reason for this incompatibility may be due 
to the data taken in different bands, effecting the number of stars used for the estimation.


\subsection{CMDs and membership probabilities} CMDs are robust tools to
determine the parameters of the open clusters. However, in order to be able
to use CMDs as tools to determine physical parameters of the clusters one
should first make sure that the stars in the diagram are actual members of
the cluster in question. Therefore, a process to determine the membership
probabilities ($P$) of each star in the field is necessary. We calculated this
membership probability of all the objects in the field of view using the
method given by \citet{Balaguer98}. This method takes into account the errors
of the stellar proper motions as well as the average cluster proper motion,
and uses the kernel estimation technique to obtain distribution of the data.

PPMXL \citep{Roeser10} and UCAC4 \citep{Zacharias13} are the
most often used catalogues of positions and proper motions. PPMXL is the largest catalogue
containing about 900 million objects reaching apparent magnitudes down to $V=20$. 
The UCAC4 is a more recent and precise catalogue containing about
113 million stars reaching magnitudes down to $R=16$. In this study we collect proper
motion components of individual stars from the PPMXL due to the magnitude limit and the
larger number of stars. In order to check the used proper motions with those from the literature, 
we calculated the differences of the absolute proper motions for all five cluster to the values 
listed in \citet{Kharchenko13}. The differences range from 1.7 to 5.4 mas yr$^{-1}$ which is well 
within the error ranges of the individual proper motions.
Consequently, we found excellent agreement with those of
the algorithm published by \citet{Javakhishvili06}. 
We calculated the mean and median of the differences of both methods which range 
between 7 and 12$\%$ for the five open clusters. This excellent agreement is likely 
due to the distance of the clusters ($d>1$ kpc) at which both algorithm easily 
find non-members because of their much larger proper motions than the cluster members.
The histogram of the differences efficiently discriminates the members of the cluster 
from the non-members. In order to identify the most likely members of each cluster, we
fitted the zero-age main sequence (ZAMS) of \citet{Sung13} for solar
metallicity to the $V \times (B-V)$ CMDs using only the main-sequence stars
with $P\geq50 \%$. Although all stars with $P\geq50 \%$ in the field of view show 
a large scatter in CMDs (orange crosses in Fig.\,5), the method is efficient in 
determination of main sequence of the clusters.
We also shifted the fitted main sequence to brighter $V$
magnitudes to take into account the effect of binary stars (see Fig.\,5).
Finally, we assumed that the stars with $P\geq50 \%$ within this band-like
region are the most likely main-sequence members of the clusters. The red dots outside 
the main-sequence band in Fig.\,5 indicate that some stars with $P\geq50 \%$ have already 
left the ZAMS. These stars correspond to the turn-off points of the clusters. We assume that 
these objects are likely members of the clusters, as well. For further analysis, we considered 
the stars identified with this procedure for each cluster. 

\begin{figure*}
\centering
\includegraphics[scale=0.25, angle=0]{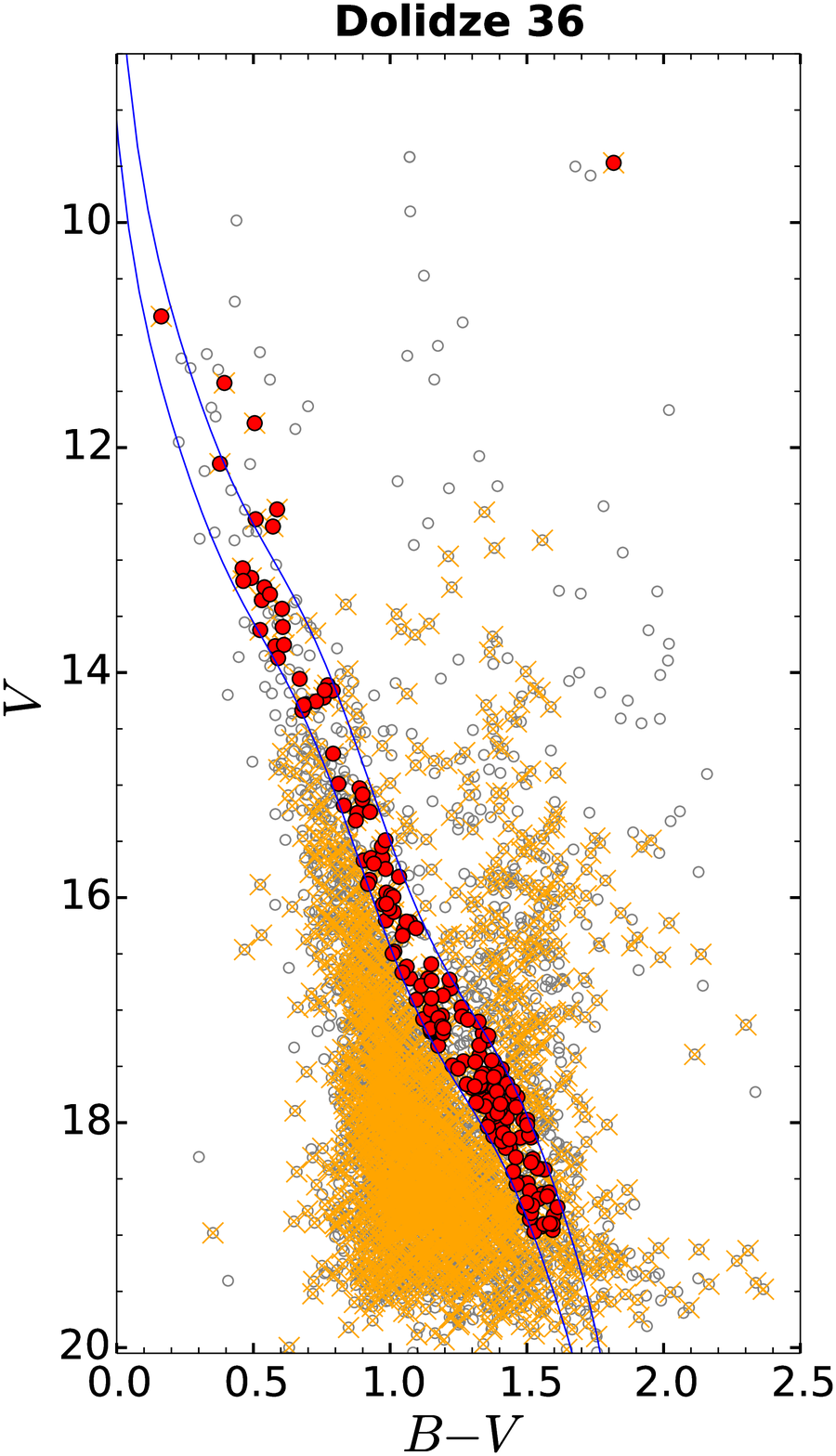}
\includegraphics[scale=0.25, angle=0]{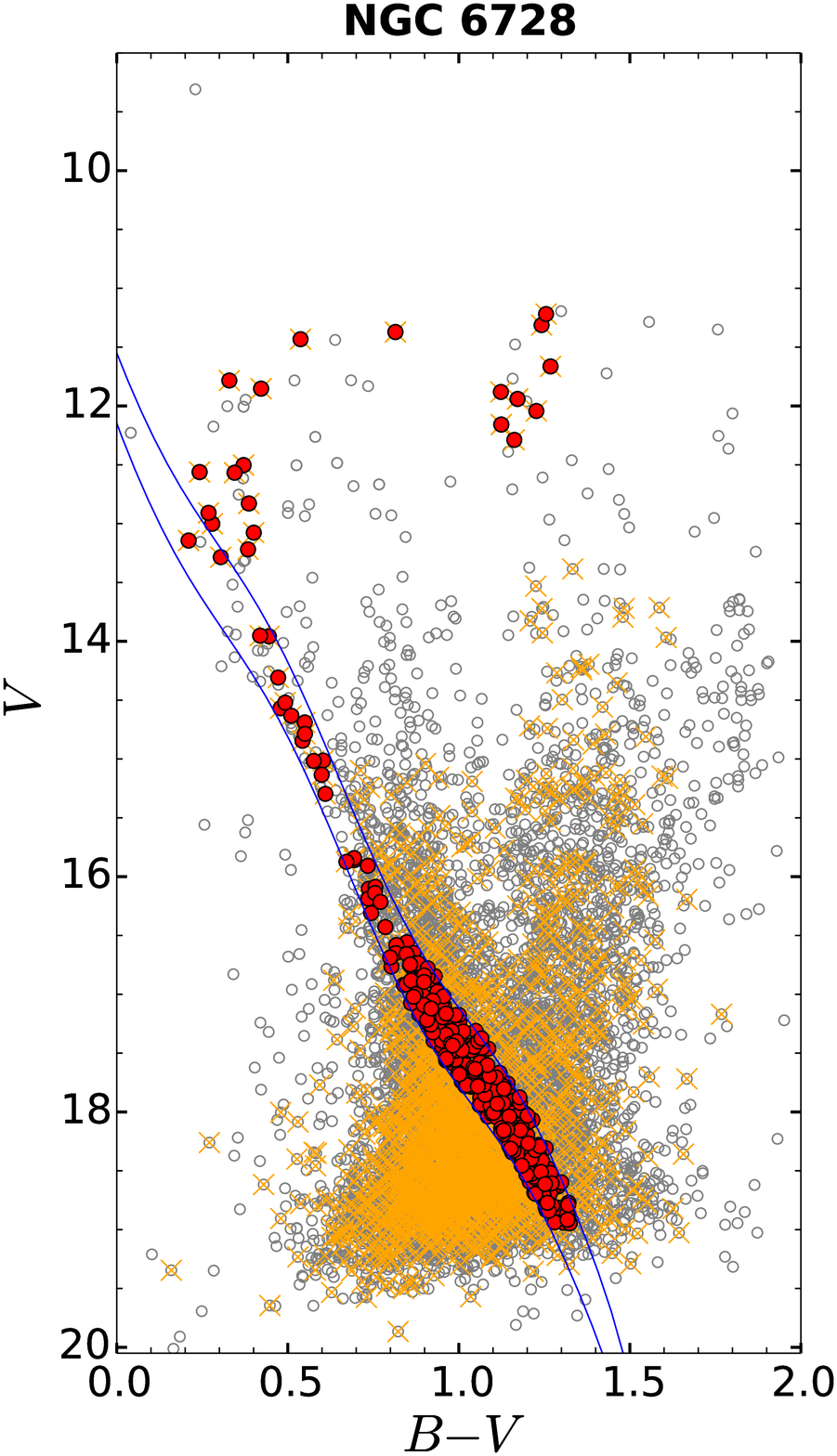}
\includegraphics[scale=0.25, angle=0]{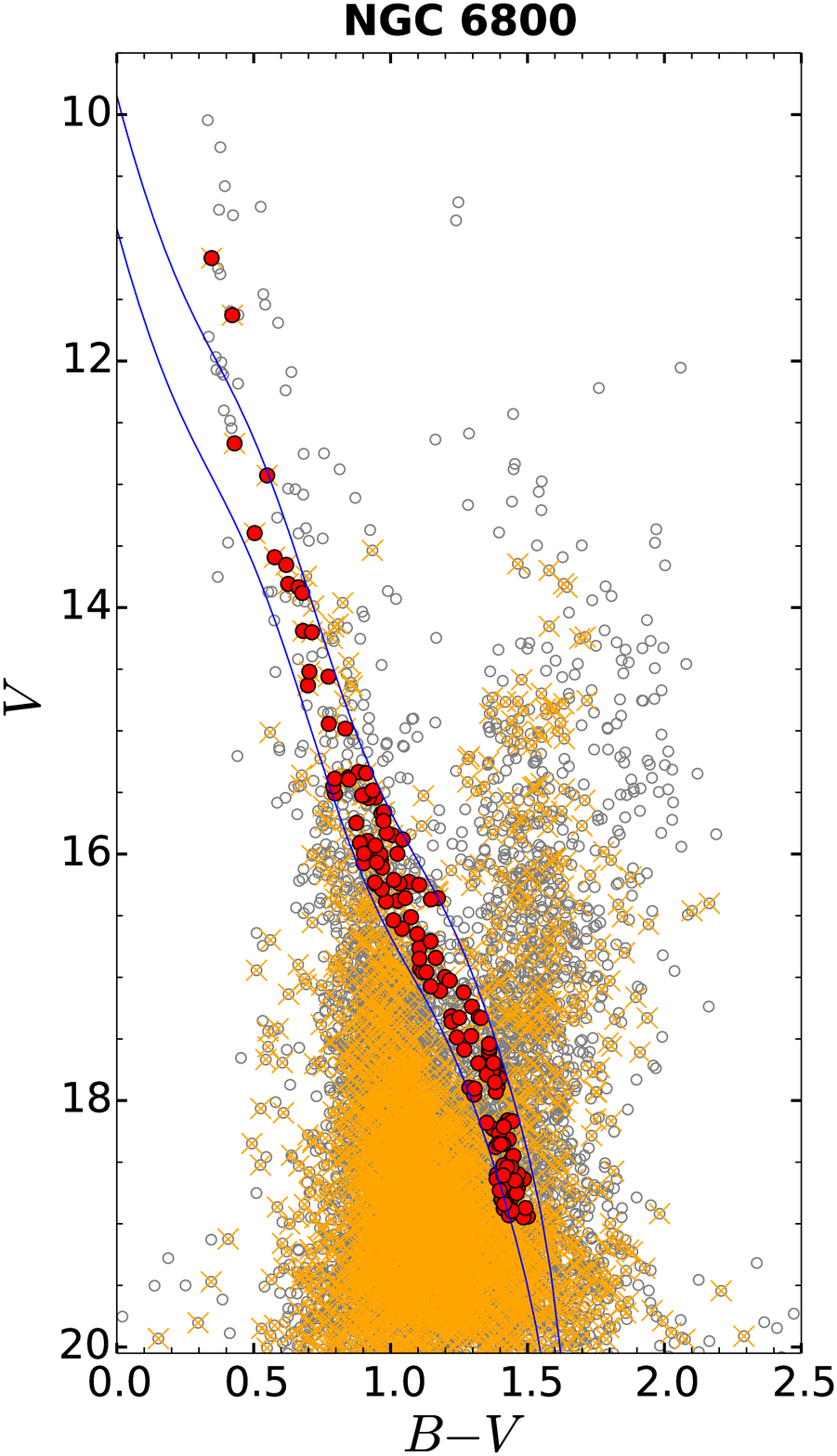}\\
\includegraphics[scale=0.25, angle=0]{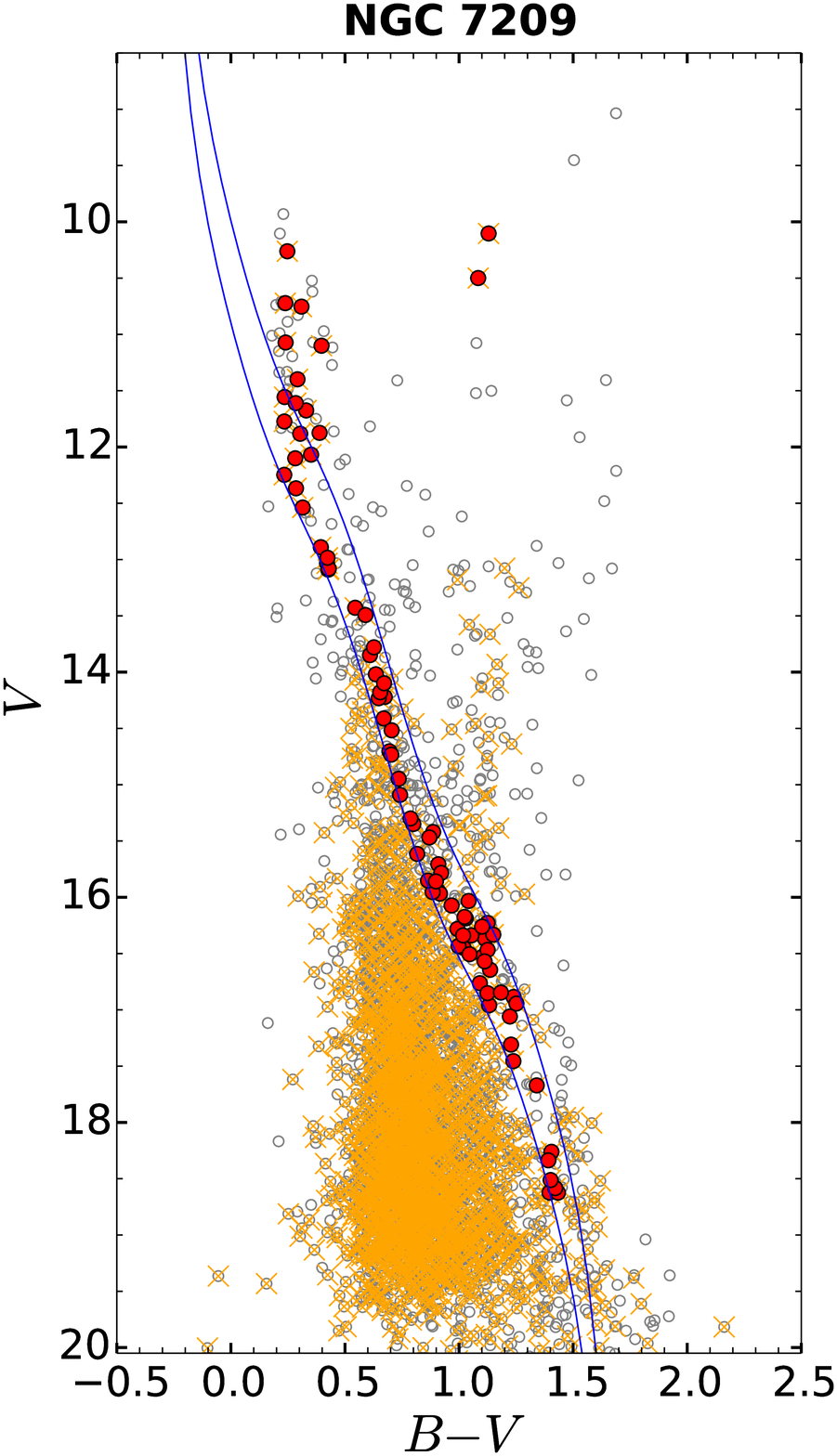}
\includegraphics[scale=0.25, angle=0]{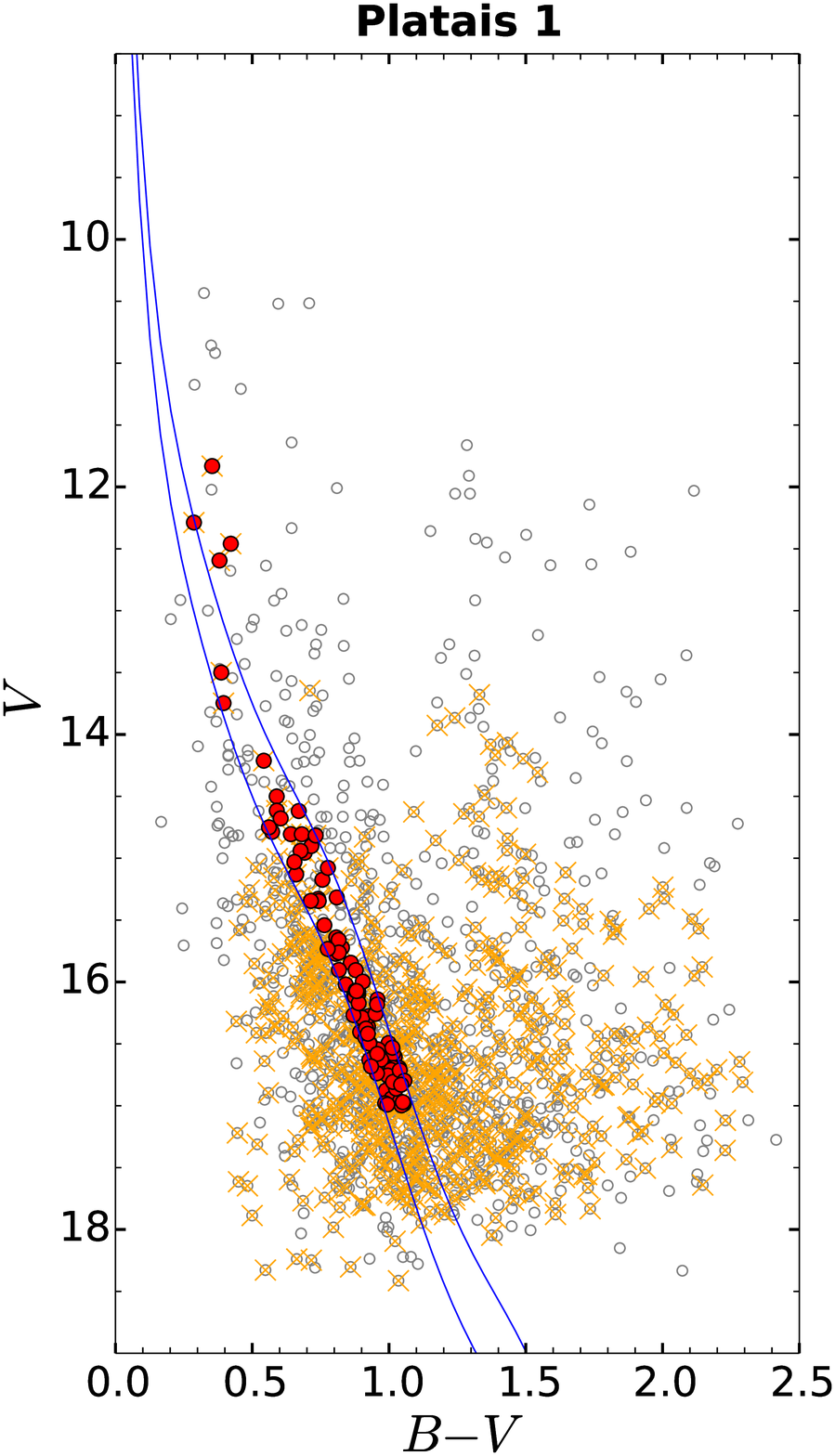}
\caption{$V\times (B-V)$ CMDs of the five clusters. Solid lines represent
the ZAMS of \citet{Sung13} and the one shifted to the brighter $V$ magnitudes
for taking into account binary stars. Red dots indicate the most probable
cluster stars within this band-like region while orange crosses represent 
the stars with $P\geq50 \%$ in the field of view. }
\end {figure*}

\begin{figure*}
\centering
\includegraphics[scale=0.35, angle=0]{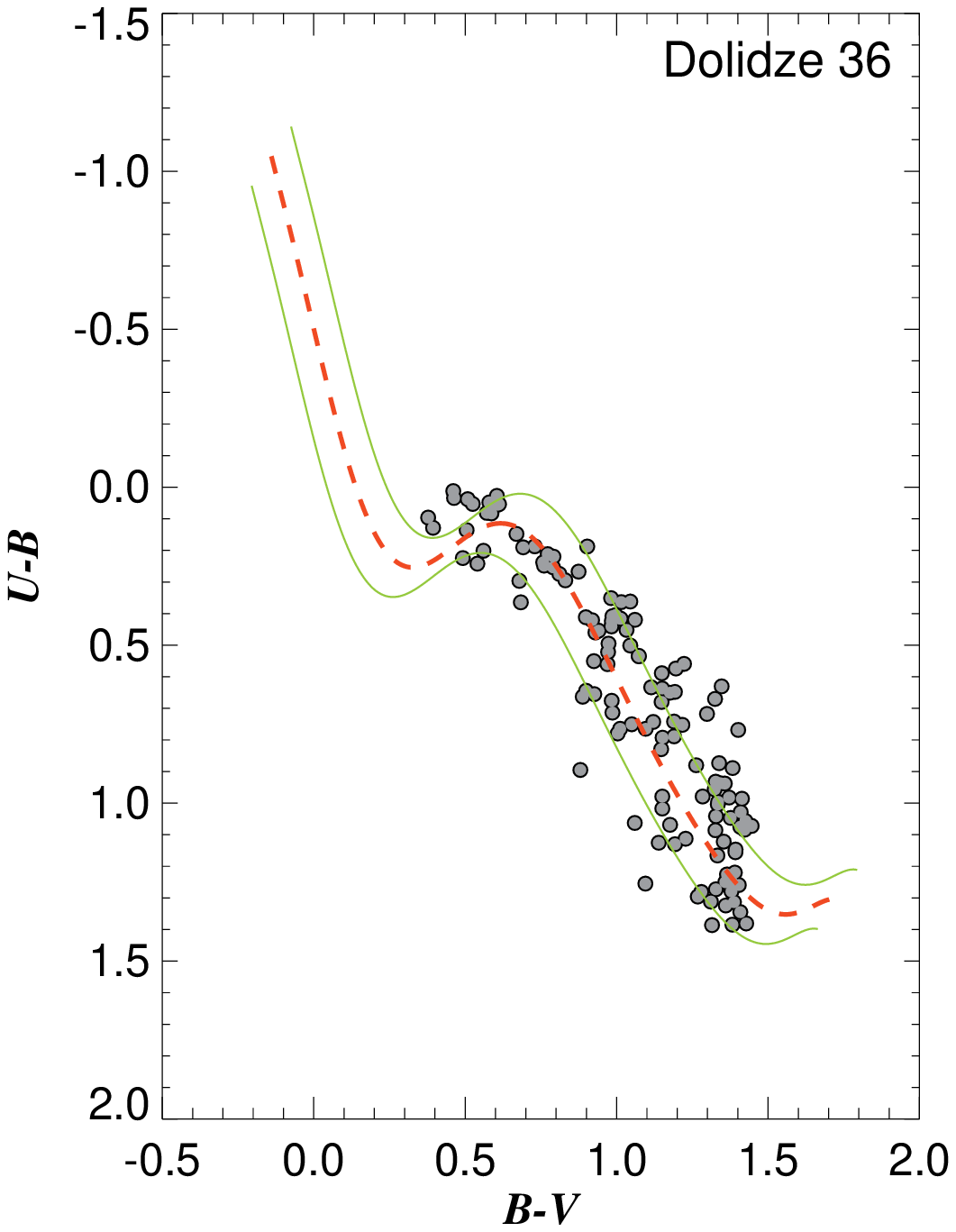}
\includegraphics[scale=0.35, angle=0]{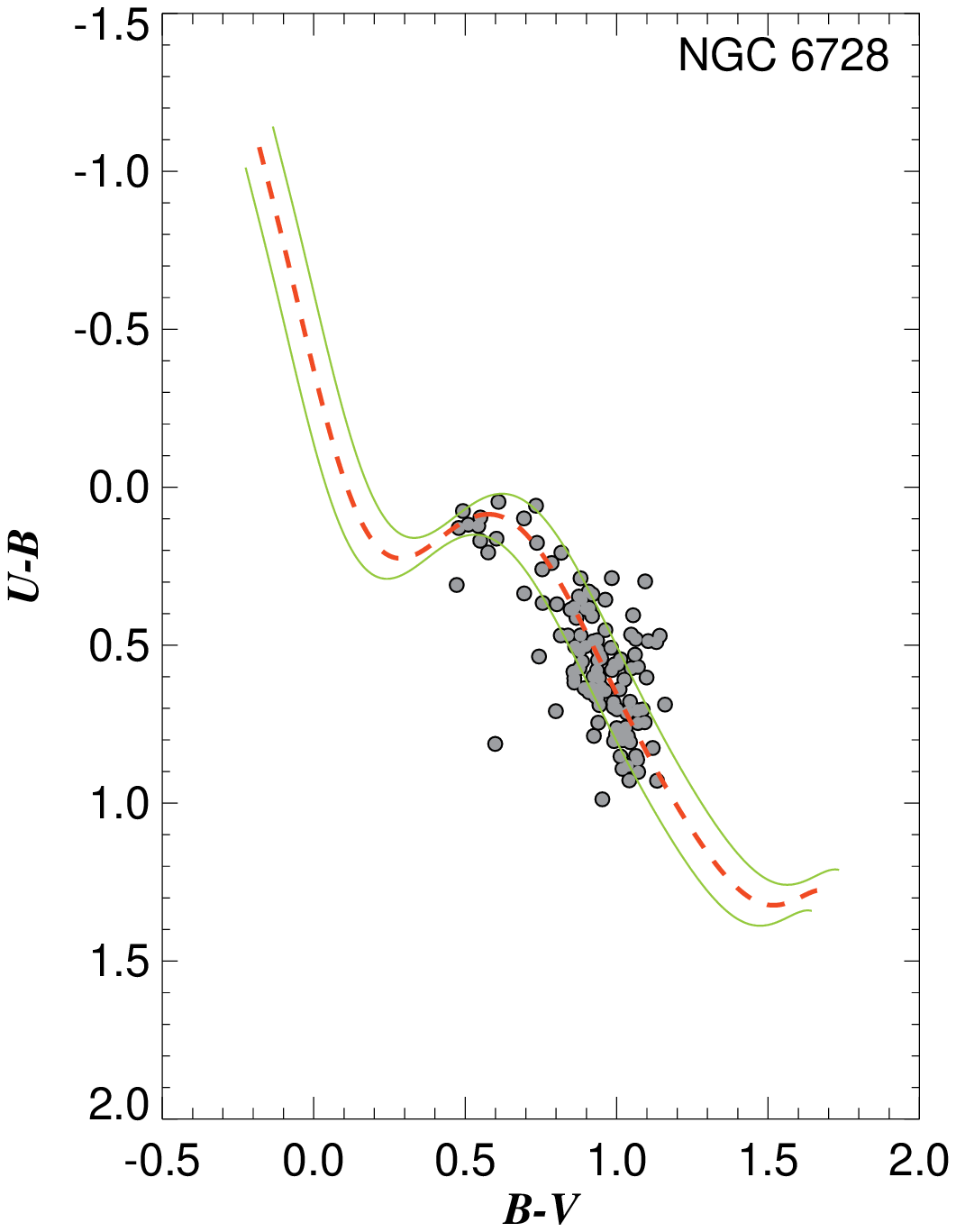}
\includegraphics[scale=0.35, angle=0]{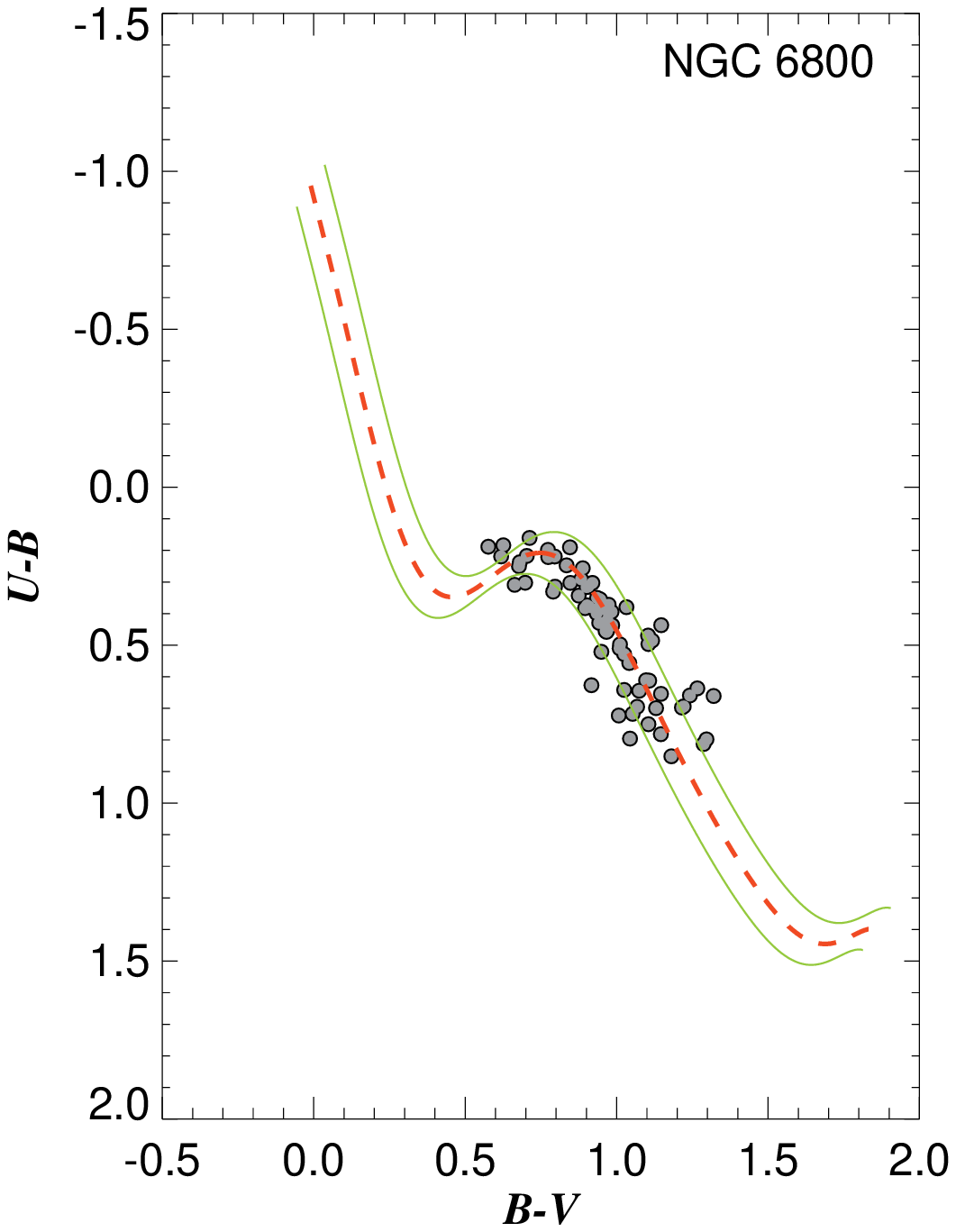}\\
\includegraphics[scale=0.35, angle=0]{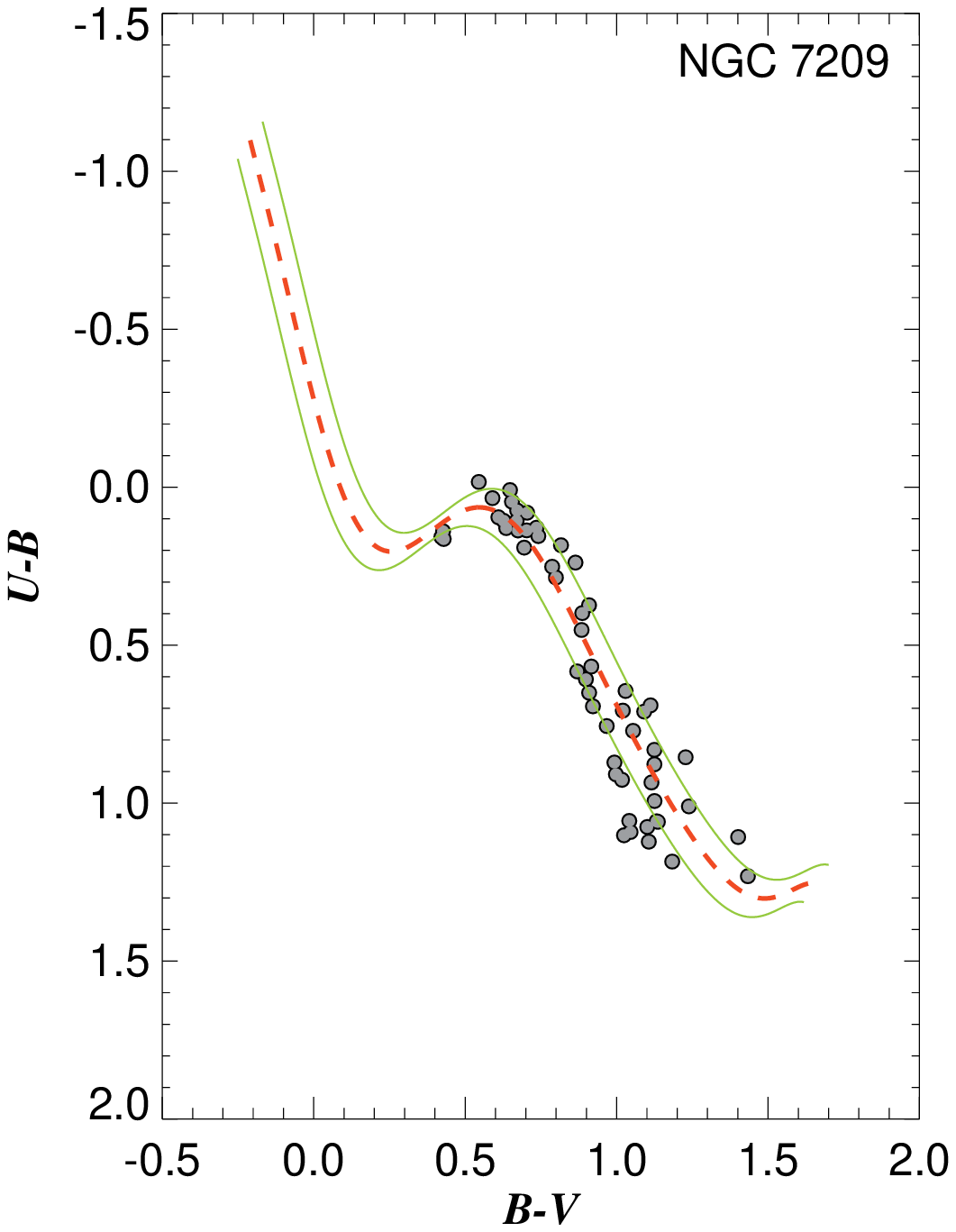}
\includegraphics[scale=0.35, angle=0]{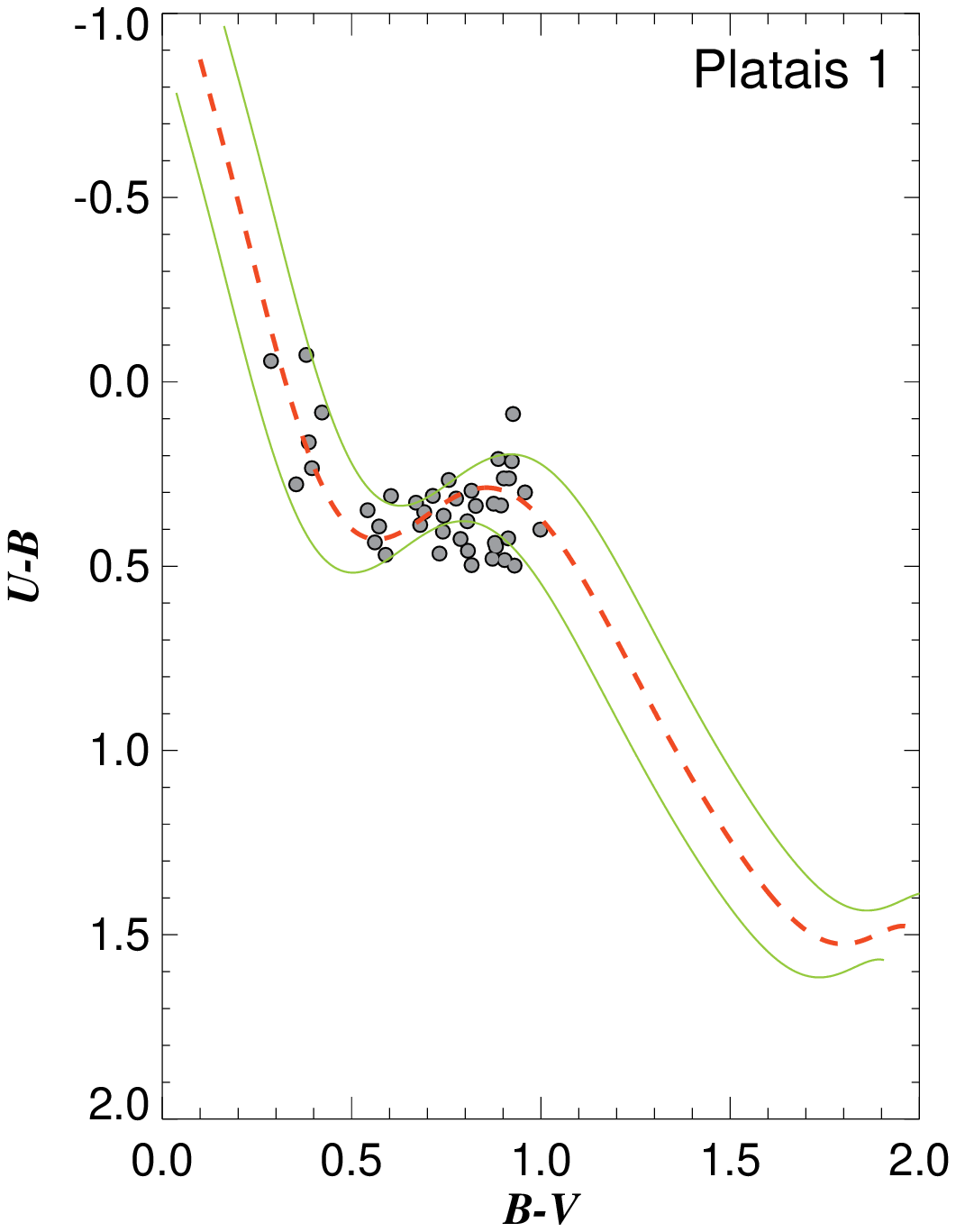}
\caption{$(U-B) \times (B-V)$ TCDs for the main-sequence stars for each
cluster. Red dashed lines are the reddened ZAMS of \citet{Sung13} fitted to
cluster stars and green solid lines represent $\pm 1\sigma$ standard
deviations, respectively.} 
\end {figure*}

\section{Astrophysical parameters of the clusters} 
The determination of astrophysical parameters by using methods 
that are independent of each other has been successfully applied in many 
of our previous studies \citep{Bilir10, Bostanci15, Ak16, Bilir16}.
In this section, astrophysical parameters of the five open clusters, such as reddening,
metallicities, distance moduli and ages, are determined via fitting models to
the observed data points selected following the method described in Section 3.3. 

\subsection{The reddening} We used the TCD constituted from the most
probable main-sequence stars, in order to derive the colour excess of each
cluster. We compared the $(U-B) \times (B-V)$ TCD of these stars with the
ZAMS of \citet{Sung13} for solar metallicity using the following equation
\citep{Garcia88}

\begin{equation}
E(U-B)=0.72\times E(B-V)+0.05\times E(B-V)^{2},
\end{equation} 
and estimated the $E(B-V)$ and $E(U-B)$ colour excesses by shifting
de-reddened ZAMS curve with steps of 0.01 mag within the range $0<E(B-V)\leq
1$ mag until the best fit is achieved with the TCD of each cluster as shown
in Fig.\,6. 
We took into account stars with $\sigma_{U-B}<0.1$ mag for 
Dolidze\,36, NGC\,6728, NGC\,6800, and NGC\,7209, and $\sigma_{U-B}<0.06$ mag for Platais\,1. 
Results are given in Table 6. The errors were calculated by
shifting the best fit curve for $\pm 1\sigma$. 

\begin{figure*}[t]
\centering
\includegraphics[scale=0.35, angle=0]{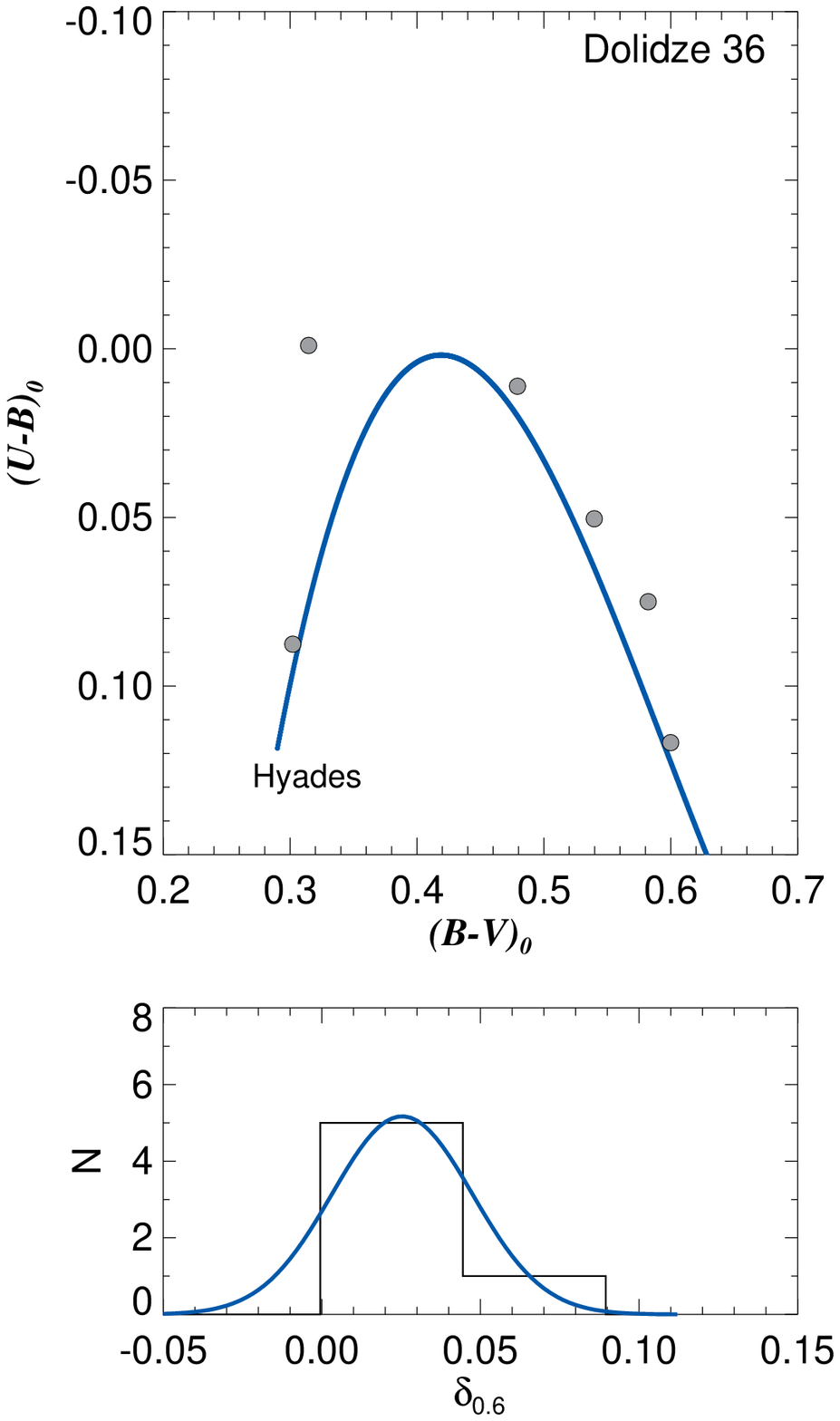}
\includegraphics[scale=0.35, angle=0]{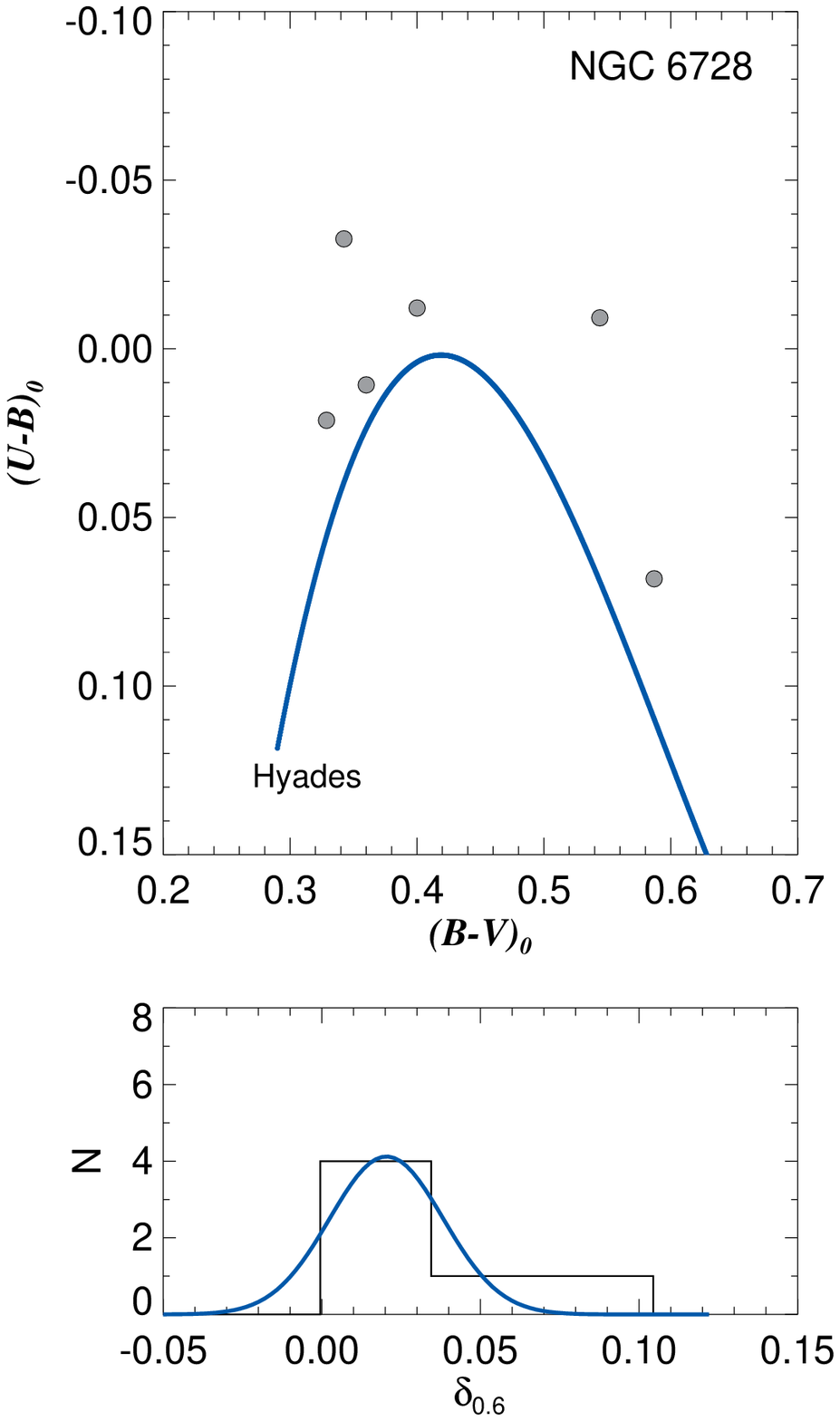}
\includegraphics[scale=0.35, angle=0]{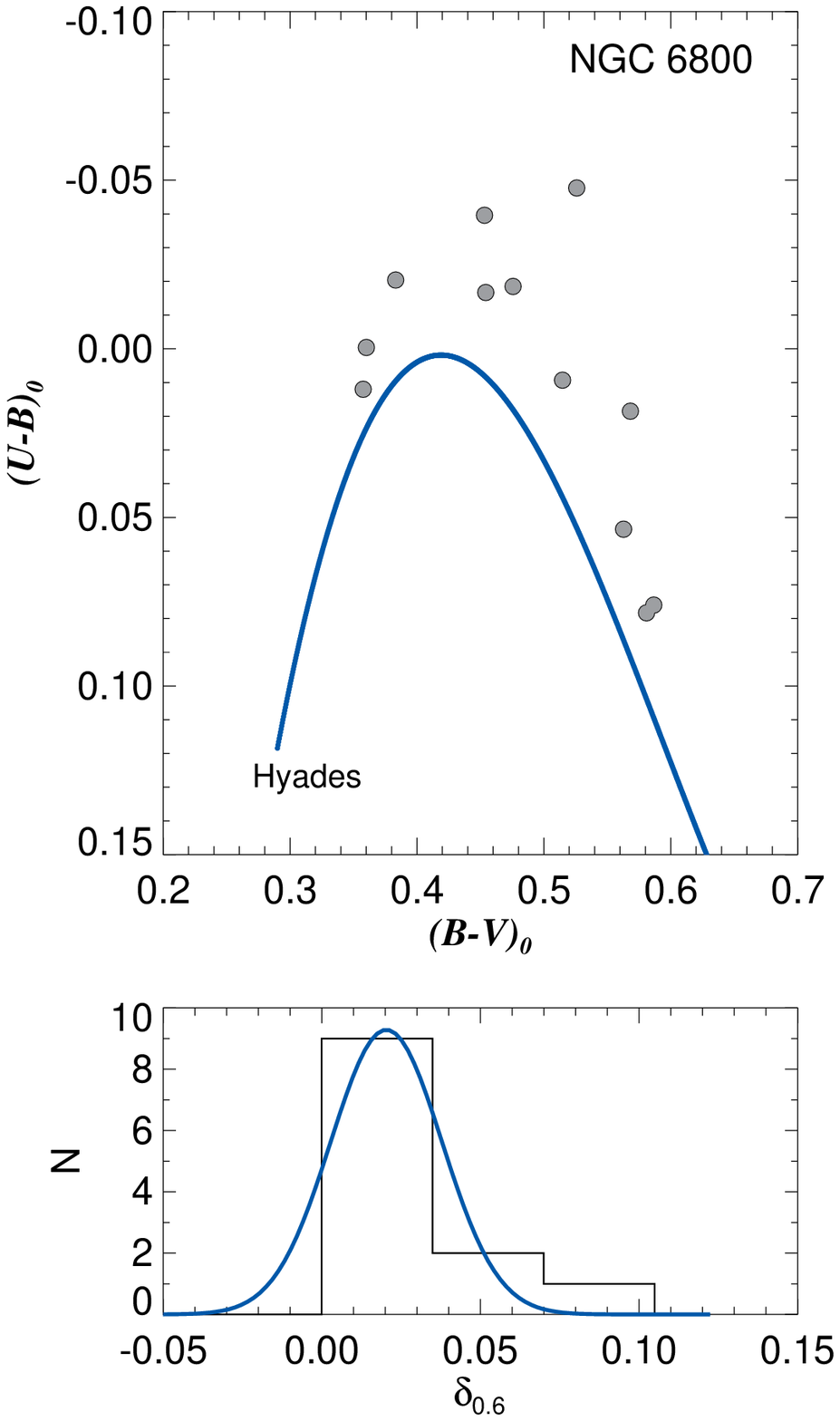}\\
\includegraphics[scale=0.35, angle=0]{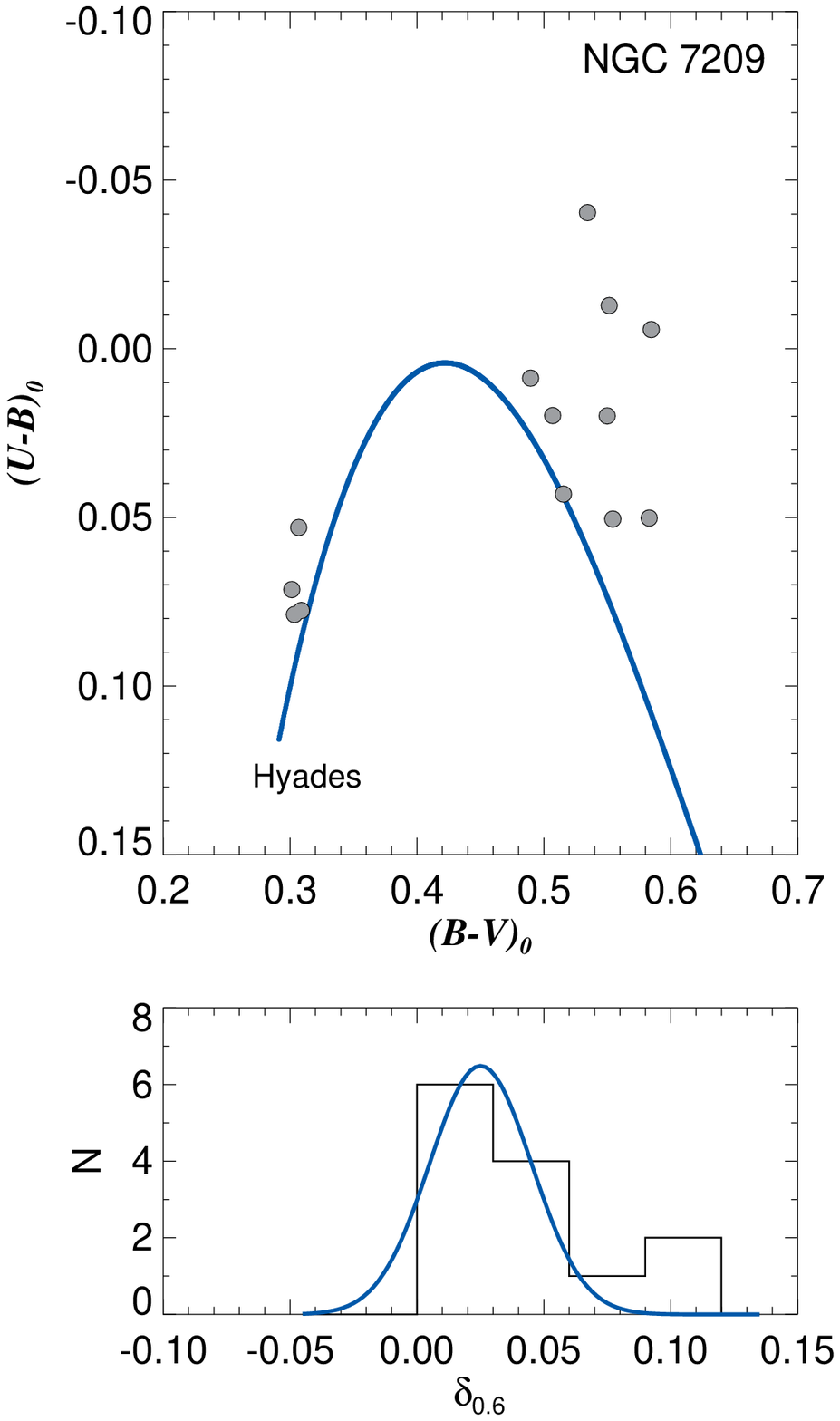}
\includegraphics[scale=0.35, angle=0]{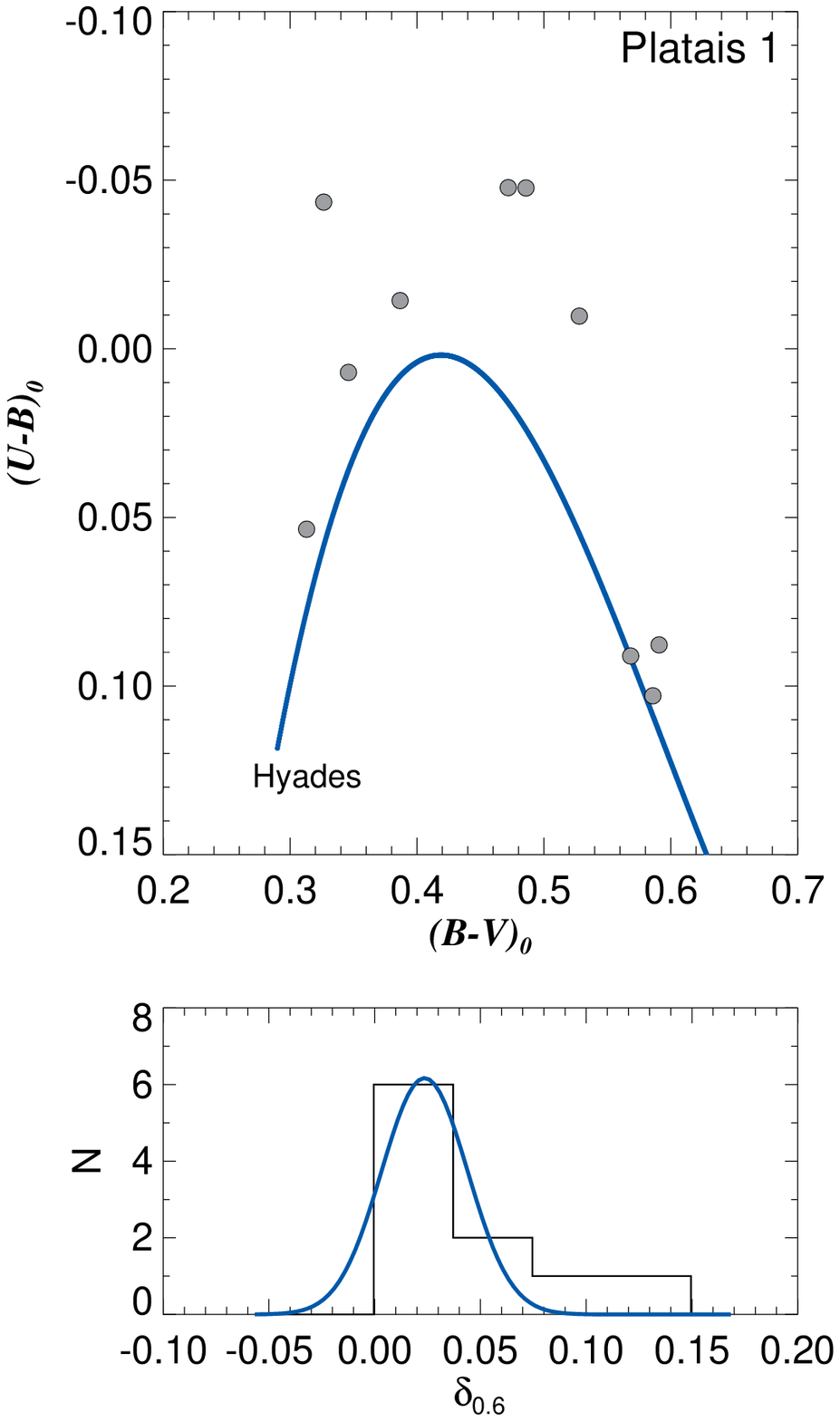}
\caption{$(U-B)_0 \times (B-V)_0$ TCDs (upper panel) and the histograms
(lower panel) for the normalized UV-excesses for main-sequence stars used for
the metallicity estimation of five open clusters. The solid lines in the
upper and lower panels represent the main-sequence of Hyades cluster and the
Gaussian fit to the histogram, respectively.} 
\end {figure*}

\begin{figure*}
\centering
\includegraphics[scale=0.20, angle=0]{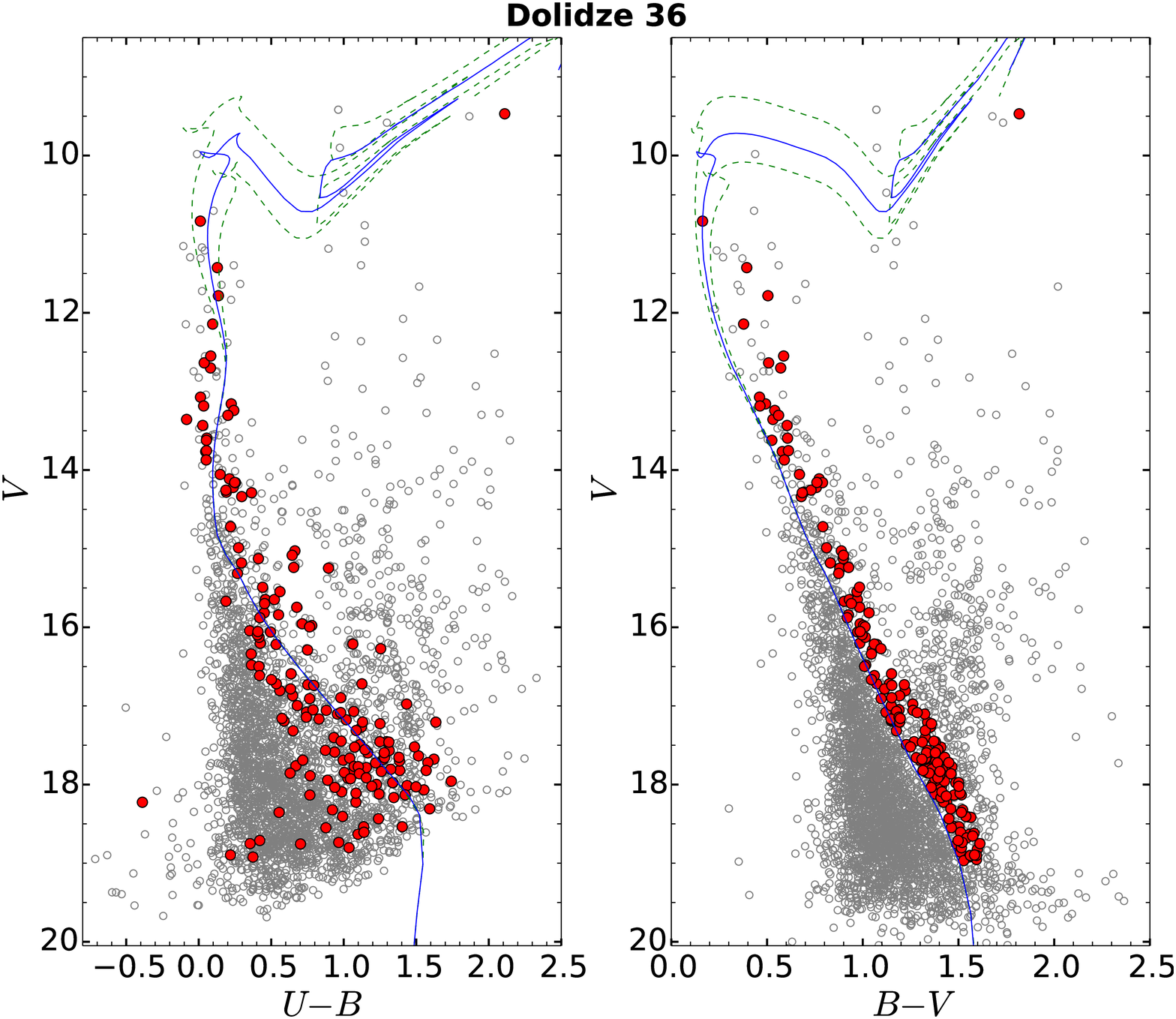}
\includegraphics[scale=0.20, angle=0]{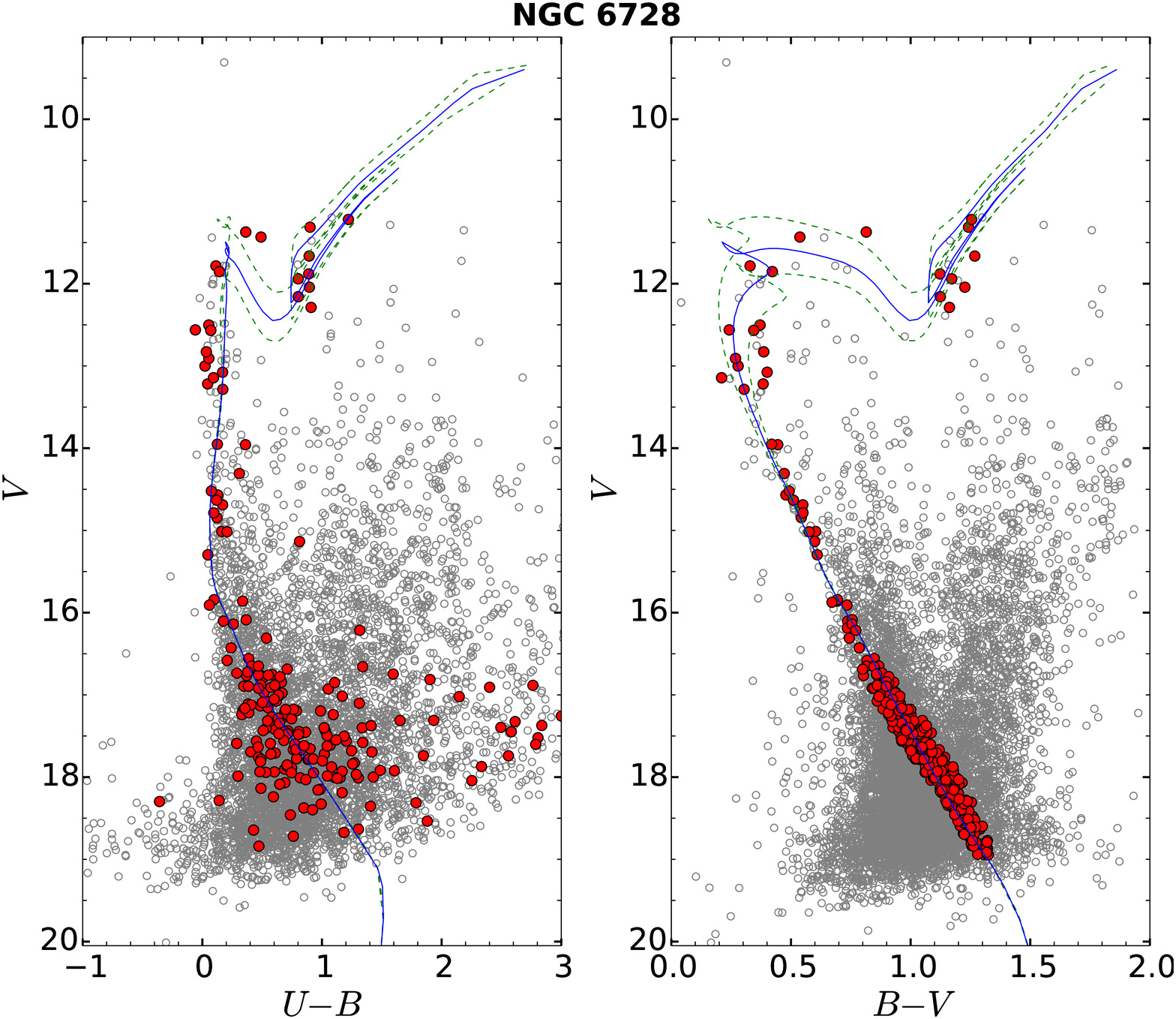}\\
\includegraphics[scale=0.20, angle=0]{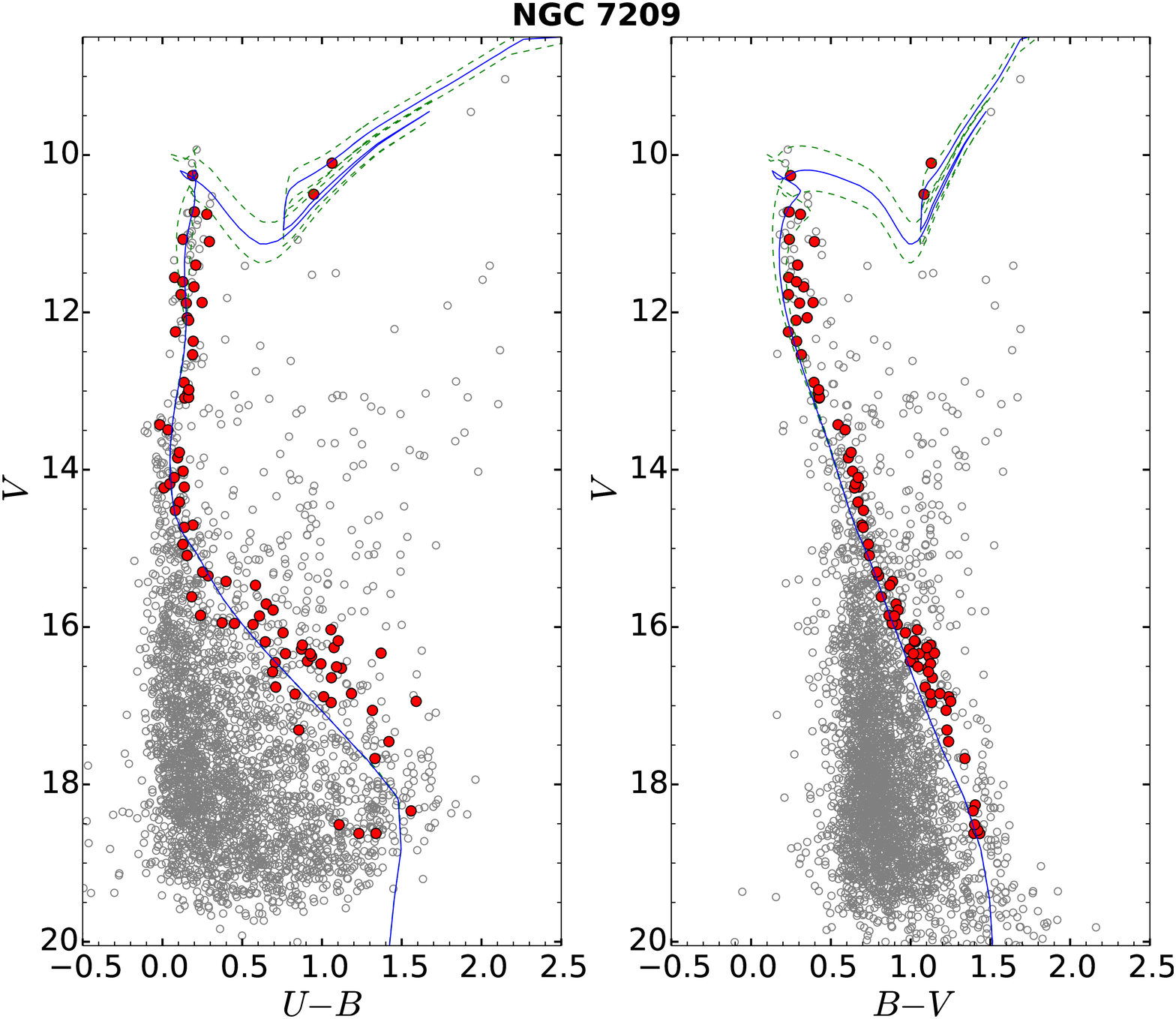}
\includegraphics[scale=0.20, angle=0]{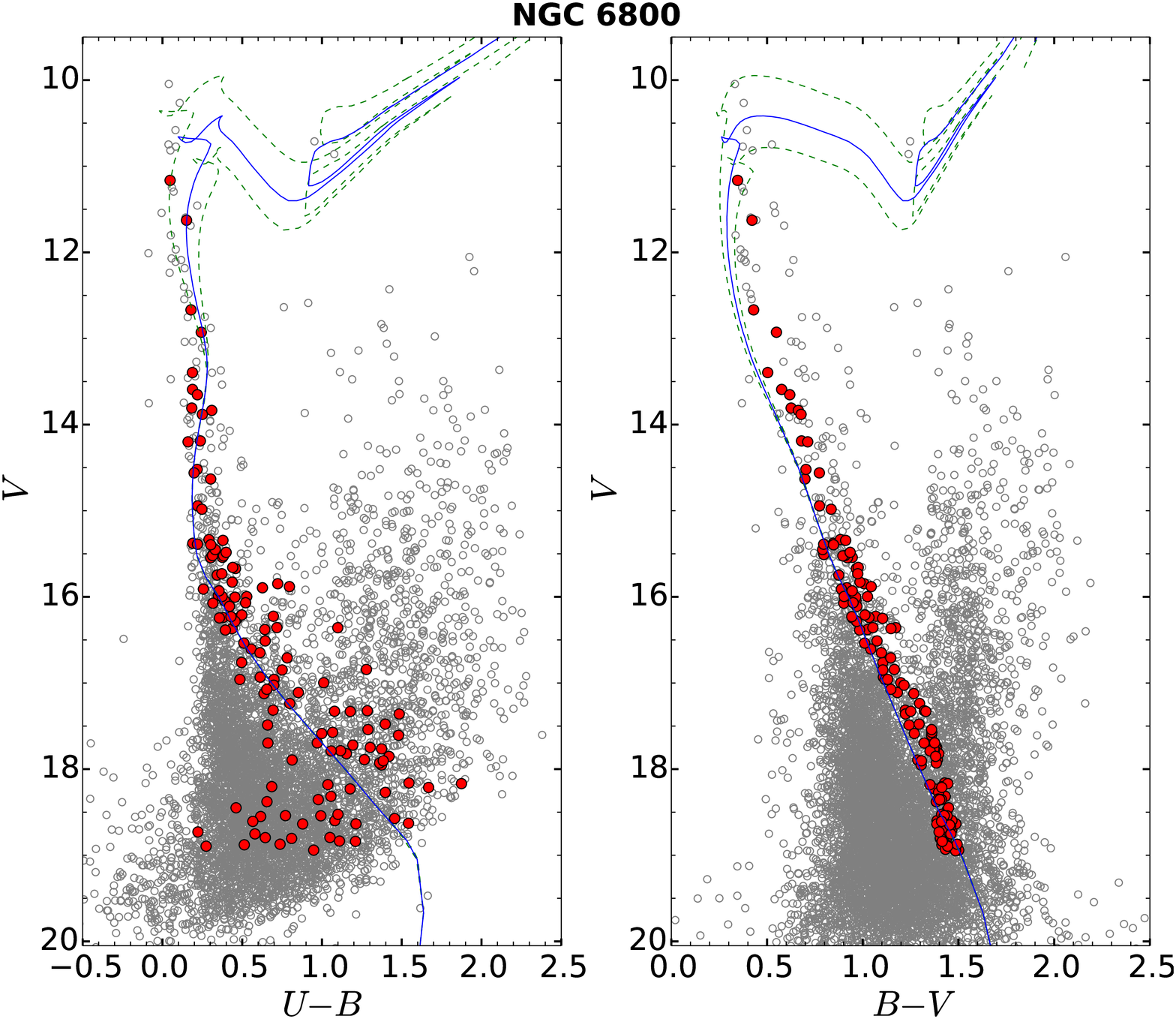}\\
\includegraphics[scale=0.20, angle=0]{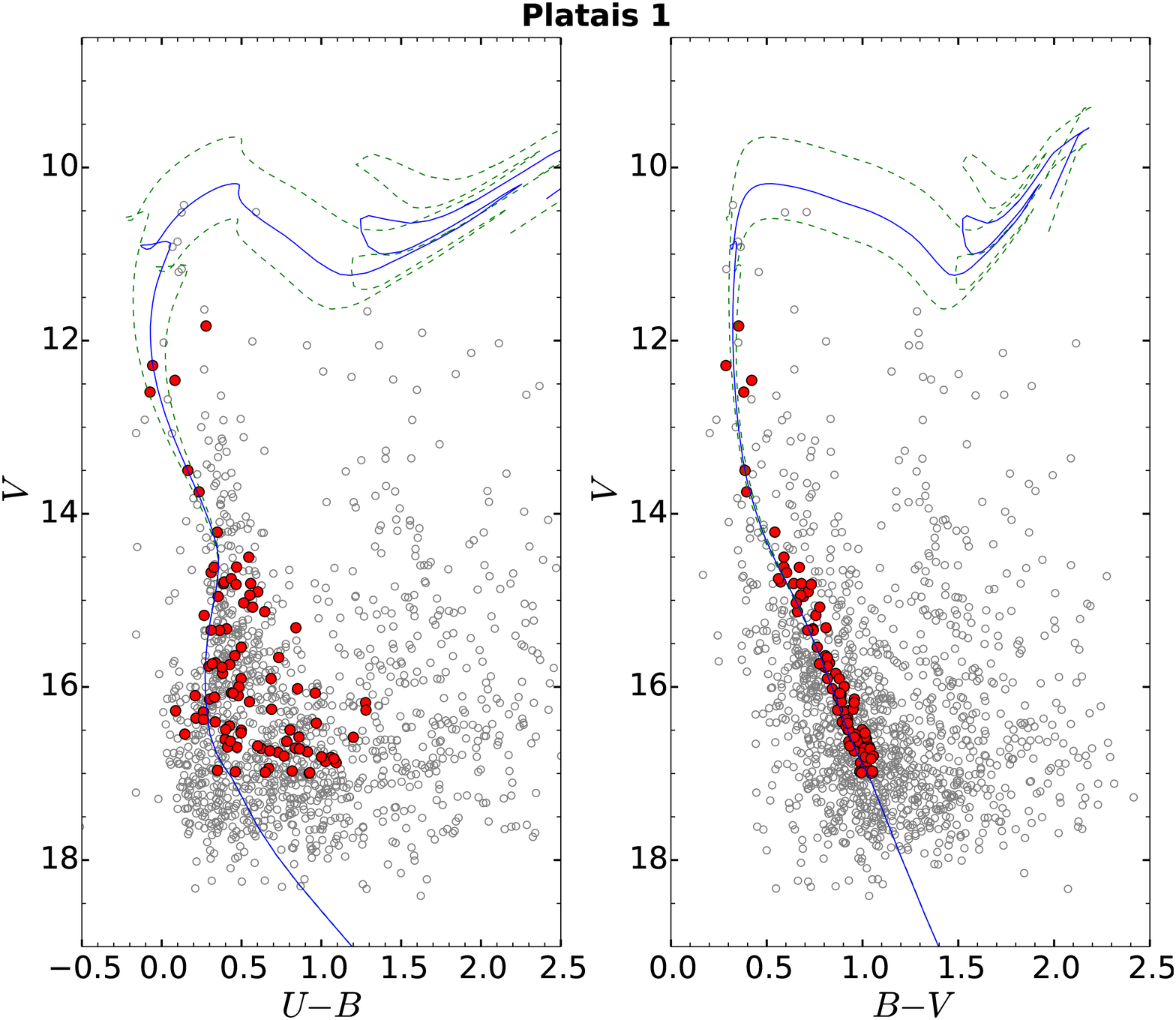}
\caption{$V \times (U-B)$ and $V \times (B-V)$ CMDs for the five open
clusters. Red circles denote the most probable members of each cluster. Blue
lines are the best fit theoretical isochrones determined in this study. The
green dashed lines represent the isochrones with estimated age plus/minus its
error.} 
\end {figure*}

\subsection{Metallicities of the clusters} We measured the photometric
metallicity of the clusters following a method based on F-G type
main-sequence stars \citep{Karaali11}. Therefore, we considered F0-G0
spectral type main-sequence stars with a membership probability $P\geq50 \%$
and with colours $0.3\leq(B-V)_0\leq0.6$ mag \citep{Cox00}. We selected six stars in Dolidze\,36, and NGC\,6728, 12 stars in NGC\,6800, 
13 stars in NGC\,7209, and 10 stars in Platais\,1.

For this method we first need to calculate normalised ultraviolet (UV)
excesses, which is the difference between a star's de-reddened $(U-B)_0$
colour index and the one corresponding to the members of the Hyades cluster
with the same de-reddened $(B-V)_0$ colour index ($\delta =
(U-B)_{0,H}-(U-B)_{0,S}$, where H and S denote to Hyades and star,
respectively). Thus, we calculated the normalised UV-excesses of member stars
for each cluster and normalised their $\delta$ differences to the UV-excess
at $(B-V)_0=0.6$ mag (i.e. $\delta_{0.6}$). The distributions of normalised 
$\delta_{0.6}$ UV-excesses were fitted with a Gaussian function and we adopt 
the Gaussian peak as result for each cluster. The $(U-B)_0\times(B-V)_0$ TCDs 
and histograms of the normalised $\delta_{0.6}$ UV-excesses of the selected 
stars are shown in Fig.\,7. The standard deviation of the metallicity distribution 
for each cluster was assumed as the uncertainty of the metallicity.
The [Fe/H] metallicity for each cluster was calculated from the following equation
\citep{Karaali11}:

\begin{equation}
{\rm [Fe/H]}=0.105-3.557\times\delta_{0.6}-14.316\times\delta_{0.6}^2.
\end{equation} 

To transform the [Fe/H] metallicities to the mass fraction $Z$ of all
elements heavier than helium, we used the following equations from Bovy who
obtained them analytically using PARSEC isochrones
\footnote{https://github.com/jobovy/isodist/blob/master/isodist/\\Isochrone.py}:   

\begin{equation}
Z_X={10^{{\rm [Fe/H]}+\log \left(\frac{Z_{\odot}}{1-0.248-2.78\times Z_{\odot}}\right)}},
\end{equation} 

and

\begin{equation}
Z=\frac{(Z_X-0.2485\times Z_X)}{(2.78\times Z_X+1)}.
\end{equation}
Here, $Z_X$ is the intermediate operation function depending on [Fe/H]
and the solar mass fraction is taken as $Z_{\odot}=0.0152$ \citep{Bressan12}.
Resulting $Z$ value for each cluster is presented in Table 6. 

\begin{table*}
\setlength{\tabcolsep}{5pt}
\begin{center}
\caption{Colour excesses ($E(B-V)$), distance moduli ($\mu_V$), distances
($d$) iron abundances ([Fe/H]) metallicities ($Z$) and ages ($t$) estimated
using two CMDs and TCDs of each cluster.}
\begin{tabular}{lcccccc}
\hline
Cluster   & $E(B-V)$      & $\mu_V$       & $d$           & [Fe/H]          & $Z$               & $t$         \\
          & (mag)         & (mag)         & (pc)          & (dex)           &                   & (Myr)       \\
\hline
Dolidze~36& 0.19$\pm$0.06 & 10.70$\pm$0.19 & 1050$\pm$90  & 0.00$\pm$0.09 & 0.0152$\pm$0.0036 & 400$\pm$100 \\
NGC~6728  & 0.15$\pm$0.05 & 11.50$\pm$0.25 & 1610$\pm$190 & 0.02$\pm$0.11 & 0.0159$\pm$0.0042 & 750$\pm$150 \\
NGC~6800  & 0.32$\pm$0.05 & 11.40$\pm$0.26 & 1210$\pm$150 & 0.03$\pm$0.07 & 0.0162$\pm$0.0027 & 400$\pm$100 \\
NGC~7209  & 0.12$\pm$0.04 & 10.50$\pm$0.18 & 1060$\pm$90  & 0.01$\pm$0.08 & 0.0154$\pm$0.0032 & 600$\pm$100 \\
Platais~1 & 0.43$\pm$0.06 & 12.50$\pm$0.29 & 1710$\pm$250 & 0.01$\pm$0.08 & 0.0154$\pm$0.0032 & 175$\pm$50  \\
\hline
\end{tabular}
\end{center}
\end{table*}

\subsection{Distance moduli and the ages of the clusters} Using the
reddening and metallicity values that we calculated above, we employed the
isochrone fitting procedure to simultaneously obtain distance moduli and ages of
the five open clusters. We shifted the theoretical isochrones provided by the
PARSEC V1.2 synthetic stellar library \citep{Bressan12, Tang14, Chen14} onto
observed $V \times (U-B)$ and $V \times (B-V)$ CMDs, respectively. Fig.\,8
shows CMDs of each cluster overplotted with the best fit theoretical
isochrones. 
In the $V \times (B-V)$ CMD of Dolidze\,36 (Fig.~8), some stars with most likely 
membership located in the $0.4<B-V<0.6 $ and $11<V<13$ magnitudes are not fitted 
well with the isochrone. The reason for this could be that these redder stars are binary stars.

We assumed the standard selective absorption coefficient as $R_V=3.1$
\citep{Schultz75} for the distance calculation. We considered the errors in 
distance moduli of the clusters for the determination of errors in the distances \citep[see also, ][]{Carraro17}. 
We listed the resulting distance moduli, distances and ages of
the clusters in Table 6.  

\subsection{Mass functions of the clusters}
Mass function (MF) indicates the relative number of stars in a unit range
of mass and denotes the rate of star formation based on stellar mass.
Considering the most likely main-sequence members of the open clusters in
this study, we first calculated $M_V$ of each star using the distance moduli
and $V$ magnitudes. We then converted $M_V$ to mass values using the best fit
theoretical PARSEC isochrones for each cluster. Fig.\,9 shows MFs of the five
open clusters. We derived the slope $x$ of mass function from the following
linear relation: $\log(dN/dM)=-(1+x)\times \log(M)+C$, where $dN$ indicates
the number of stars in a mass bin $dM$ with central mass of $M$, and $C$ is a
constant. We give the resulting values for the slopes of MFs for each cluster
in Table 7. 

\begin{table}
\setlength{\tabcolsep}{3pt}
\begin{center}
\caption{The slopes of the mass functions of clusters.}
\begin{tabular}{lccc}
\hline
Cluster    &  $N$ &  $x$ & Mass range\\
\hline
Dolidze~36 & 122 & -1.39$\pm$0.91 & $0.7<M/M_{\odot}<1.7$ \\
NGC~6728   & 182 & -1.58$\pm$0.61 & $0.7<M/M_{\odot}<1.9$ \\
NGC~6800   & 110 & -1.58$\pm$0.50 & $0.7<M/M_{\odot}<1.9$ \\
NCG~7209   &  60 & -1.31$\pm$0.35 & $0.7<M/M_{\odot}<1.8$ \\
Platais~1  &  40 & -1.49$\pm$0.59 & $1.3<M/M_{\odot}<2.1$ \\
\hline
\end{tabular}
\end{center}
\end{table}

\begin{figure}
\centering
\includegraphics[width=8.075cm,height=14.45cm, angle=0]{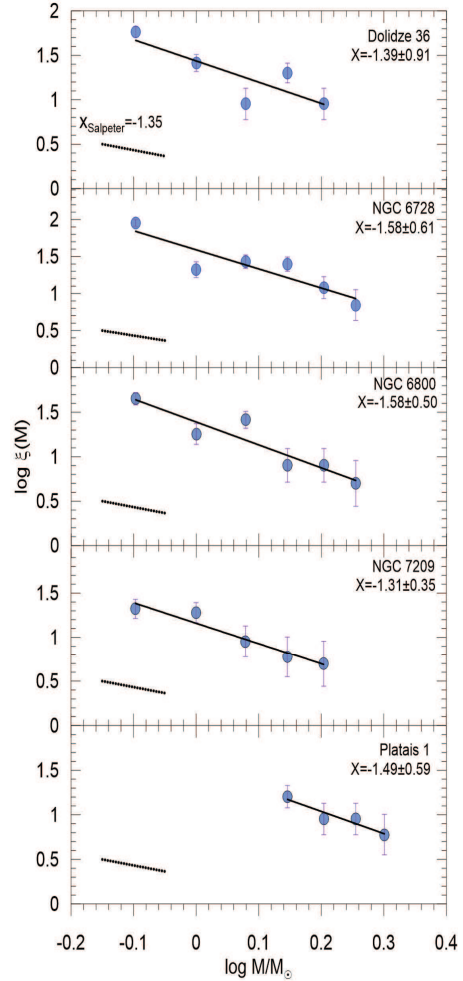}
\caption{Mass functions of clusters determined from the stars with the membership probability  $P\geq50\%$.} 
\end {figure}

\subsection{Integrated magnitudes and colours of the clusters}
We used the following equation to convert apparent magnitudes and
absolute magnitudes in {\it UBV} passbands to flux values for the stars with
the membership probability $P>0\%$ in order to not to miss member stars of open clusters and then summed flux of all these stars to obtain $U$, $B$, $V$, and $M_V$ integrated magnitudes ($I(m)$) of the clusters:  

\begin{equation}
I(m)=-2.5 \times \log \Biggr[ \sum_i(10^{-0.4\times m_i})\Biggr]. 
\end{equation}
From reddening and distance moduli values (see Table 6), we calculated the
integrated magnitude, colours, and absolute magnitude of each cluster and
listed them in Table 8.

\cite{Lata02} found relations between integrated parameters and ages
using 352 open clusters. We also calculated the integrated $I(B-V)_0$ and
$I(M_V)$ values with their relations given below:

\begin{equation}
I(B-V)_0=1.99-0.79\times (\log t)+0.07\times(\log t)^2, 
\end{equation}

\begin{equation}
I(M_V)=-36.53+6.90\times (\log t)-0.36\times(\log t)^2, 
\end{equation}
where $t$ indicates the age of the cluster. All in all, when we compare our
integrated colour and absolute magnitudes calculated from the stars located
in the direction of clusters with those calculated using Eqs. (6) and (7)
from \citet{Lata02}, we see that they are in good agreement except Dolidze~36, 
as seen in Table 8. When we look over the large samples of
\citet{Lata02}, a few of them have similar colour and absolute magnitude
values with Dolidze~36. These values are in the scattered region of Fig.\,4
given by \citet{Lata02}, since Dolidze~36 has fainter and redder magnitudes
than the other clusters. The most likely reason is that, in a cluster, as the
massive stars evolve the dynamical evolution produces mass segregation
\citep{Nilakshi02}. 

\begin{table*}
\setlength{\tabcolsep}{5pt}
\begin{center}
\caption{Integrated absolute magnitudes and colours of the clusters.}
\begin{tabular}{lcccccccc}
\hline
      & \multicolumn{4}{c}{This study} & \multicolumn{2}{c}{\citet{Lata02}} & \multicolumn{2}{c}{Differences}\\
\hline
Cluster & $I(V_0)$ & $I(U-B)_0$ & $I(B-V)_0$ & $I(M_V)$ & $I(B-V)_0$ & $I(M_V)$ & $\Delta I(B-V)_0$ & $\Delta I(M_V)$\\
\hline
Dolidze~36 & 6.713 &0.279  & 0.811 & -3.398 & 0.374 & -3.814 & 0.437 &  0.416\\
NGC~6728   & 7.337 &0.220  & 0.511 & -3.699 & 0.492 & -3.648 & 0.019 & -0.051\\
NGC~6800   & 6.588 &-0.037 & 0.320 & -3.819 & 0.374 & -3.814 & -0.054& -0.005\\
NGC~7209   & 6.470 & 0.177 & 0.362 & -3.657 & 0.449 & -3.701 & -0.087&  0.044\\
Platais~1  & 7.090 & 0.037 & 0.329 & -4.077 & 0.234 & -4.114 &  0.095&  0.037\\
\hline
\end{tabular}
\end{center}
\end{table*}

\section{Discussion and Conclusions}
This work presents the fundamental parameters of five open clusters,
namely Dolidze 36, NGC 6728, NGC 6800, NGC 7209, and Platais 1, obtained from
CCD {\it UBV} observations, which were analysed in detail.
Astrometric data were used to calculate the membership probabilities of the
stars in the field of view of each cluster. We took into account the most
probable member stars of the clusters to determine precisely the
astrophysical parameters of each cluster.

Simultaneous determination of astrophysical parameters by fitting the
theoretical stellar evolutionary isochrones to the observed CMDs can suffer
from the reddening-age degeneracy \citep{Anders04, King05, Bridzius08,
deMeulenaer13, Janes14}. Therefore, we independently found the parameters of
the clusters. The distances and ages of the clusters were derived
by fitting TCDs and CMDs with the theoretical PARSEC models \citep{Bressan12}
while metallicities of the clusters were obtained using the method given by
\citet{Karaali11}, based on F0-G0 spectral type main-sequence stars
\citep{Cox00}. This strategy allows us to reduce in part the effect of the
reddening-age degeneracy on the parameters. Results of the cluster parameters
are given in Table 6. As specified before, the clusters in our sample have
only a few studies in the literature. Previous results are listed in Table 1.

\subsection{Dolidze~36} The reddening, the distance modula, the distance,
the metallicity and the age of Dolidze~36 were obtained as
$E(B-V)=0.19\pm0.06$ mag, $\mu_V=10.70\pm0.19$ mag, $d=1050\pm90$ pc,
[Fe/H]=+0.00$\pm$0.09 dex, and $t=400\pm100$ Myr, respectively. Dolidze\,36 is
one of the clusters poorly studied in literature. Our reddening value agrees
within the quoted errors with the value in the catalogue given by
\citet{Kharchenko05}. However, our distance is larger and age is somewhat younger
when comparing those from \citet{Kharchenko05}.

\subsection{NGC~6728} The reddening, the distance modula, the distance,
the metallicity and the age of NGC\,6728 were determined as
$E(B-V)=0.15\pm0.05$ mag, $\mu_V=11.50\pm0.25$ mag, $d=1610\pm190$ pc,
[Fe/H]=+0.02$\pm$0.11 dex, and $t=750\pm150$ Myr, respectively. There is
only one previous study for the cluster by \citet{Kharchenko05}, 
who determined cluster parameters using the isochrones with
solar metallicity ($Z=0.019$) in their catalogues. Our values for reddening
and age agree within the quoted errors, but the distance modula and the
distance from this study are slightly larger than their results (see Table 1).

\subsection{NGC~6800} The reddening, the distance modula, the distance,
the metallicity and the age of NGC\,6800 are $E(B-V)=0.32\pm0.05$ mag,
$\mu_V=11.40\pm0.26$ mag, $d=1210\pm150$ pc, [Fe/H]=+0.03$\pm$0.07 dex, and
$t=400\pm100$ Myr, respectively. \citet{Ananjevskaja15} studied NGC\,6800
comprehensively using photographic plates from the Pulkovo ``Fantasy''
automated measuring system and point source catalogue \citep{Cutri03} of Two
Micron All Sky Survey \citep[2MASS,][]{Skrutskie06}. They obtained the
parameters of the cluster from 109 member stars in $V \times (B-V)$ and 
$J\times (J-K_S)$ CMDs. Our results do not seem to be in agreement with their
results. Their reddening value is slightly larger then ours while the distance
modula, distance and age are smaller then our results. The difference could
be attributed to the quality of their data (photographic plates) and the
method they used. \citet{Kharchenko05} gave the parameters of
the cluster in their catalogue, as well. Only the age of the cluster is in
good agreement with our value.

\subsection{NGC~7209} The reddening, the distance modula, the distance,
the metallicity and the age of NGC\,7209 were found as $E(B-V)=0.12\pm0.04$
mag, $\mu_V=10.50\pm0.18$ mag, $d=1060\pm90$ pc, [Fe/H]=+0.01$\pm$0.08 dex,
and $t=600\pm100$ Myr, respectively. NGC\,7209 has a number of previous
studies \citep{Kharchenko05, Vansevicius97, Malysheva97, Twarog97, Lynga87,Claria96,
Piatti95}. We summarized early works in Table 1. Our reddening value is in
agreement within the quoted errors with those of \citet{Kharchenko05}, \citet{Lynga87}
\citet{Claria96}, and \citet{Piatti95}, but somewhat smaller than the value
reported by {\citet{Vansevicius97}, \citet{Malysheva97}, and
\citet{Twarog97}. Distance modula and distance in our study agree with the
values from previous studies. On the other hand, we found the age of the
cluster older than those in the literature. Our metallicity value agrees
within the quoted errors with the value given by \citet{Twarog97}, but it is
larger than those of \citet{Vansevicius97} and \citet{Piatti95}.

\subsection{Platais~1}

The reddening, the distance modula, the distance, the metallicity and the
age of Platais\,1 are $E(B-V)=0.43\pm0.06$ mag, $\mu_V=12.50\pm0.29$ mag,
$d=1710\pm250$ pc, [Fe/H]=+0.01$\pm$0.08 dex, and $t=175\pm50$ Myr,
respectively. This is the youngest cluster with the largest reddening value at
the farthest distance in our sample. There is only one photometric study
on the cluster \citep{Turner94}. The astrophysical parameters of the cluster in 
our study are well consistent with those from \citet{Turner94} while the structural 
parameters are slightly in agreement.

Our sample consists of young open clusters (a mean value of $\sim$ 500
Myr), which corresponds to an average time scale for dynamics to have not}
influence upon the IMF of the stellar systems \citep{Sagar98, Sagar01}.
Therefore, we can assume that MF of the clusters could be equivalent to their
IMF. We derived the slopes of the mass functions for the member stars of each
cluster, which have mass values ranging from 0.7 to 2.1 $M/M_{\odot}$. The
MF slopes of the five open clusters vary between 1.31 and 1.58 (see Table 7),
which are comparable to the value of 1.35 reported by \citet{Salpeter55}
within the quoted errors. Additionally, our calculations for integrated
absolute magnitudes and colours are in agreement with those derived using
relations of \citet{Lata02}.

\section{Acknowledgments} 
We thank the anonymous referee for his/her insightful and constructive 
suggestions, which significantly improved the paper.
This study has been supported in part by the Scientific and Technological Research 
Council (T\"UB\.ITAK) 113F270. Part of this work was supported by the Research Fund 
of the University of Istanbul, Project Numbers: FDK-2016-22543 and BYP48482. We thank to 
T\"UB\.ITAK for a partial support in using T100 telescope with project 
number 15AT100-738. We also thank to the on-duty observers and members of the 
technical staff at the T\"UB\.ITAK National Observatory for their support before 
and during the observations.

\end{document}